\documentclass{aa}
\usepackage{graphicx,natbib,psfrag,amssymb,amsmath}
\usepackage{txfonts}
\bibpunct{(}{)}{;}{a}{}{,}

\begin{document}

\title{The excitation of water in the S140 photon dominated region.}
\author{Dieter R.~Poelman
           \inst{1,2}
           \and
        Marco Spaans
           \inst{1}
        }
\offprints{D.R.Poelman@astro.rug.nl}
\institute{Kapteyn Astronomical Institute, P.O. Box 800, 9700 AV Groningen, the Netherlands\\
 \email{D.R.Poelman@astro.rug.nl}
        \and
         SRON Netherlands Institute for Space Research, Landleven 12, 9747 AD Groningen, the Netherlands
           }

   \date{Received / Accepted}

\abstract{
We consider the excitation of water in the Photon Dominated Region (PDR) \object{S140}. With the use of a three-dimensional escape probability method we compute the level populations of ortho- and para-\element[][]{H_2O} up to $\sim$ 350 \mbox{K} (i.e., 8 levels), as well as line intensities for various transitions. Homogeneous and inhomogeneous models are presented with densities of $10^4$--$10^5$ \mbox{$\mathrm{cm^{-3}}$} and the differences between the resulting intensities are displayed. Density, temperature, and abundance distributions inside the cloud are computed with the use of a self-consistent physi-chemical (in)homogeneous model in order to reproduce the line intensities observed with SWAS, and to make predictions for various lines that HIFI will probe in the future.\\
Line intensities vary from $\sim$ $10^{-13}$ \mbox{$\mathrm{erg}\ \mathrm{cm}^{-2}\ \mathrm{s}^{-1}\ \mathrm{sr}^{-1}$} to a few times $10^{-6}$ \mbox{$\mathrm{erg}\ \mathrm{cm}^{-2}\ \mathrm{s}^{-1}\ \mathrm{sr}^{-1}$}. We can reproduce the intensity for the $\mathrm{1_{10}}$ $\rightarrow$ $\mathrm{1_{01}}$ line observed by the SWAS satellite. It is found that the $\mathrm{2_{12}}$ $\rightarrow$ $\mathrm{1_{01}}$ line is the strongest, whereas the $\mathrm{3_{12}}$ $\rightarrow$ $\mathrm{2_{21}}$ line is the weakest, in all the models. It is found that the $\mathrm{1_{10}}$ $\rightarrow$ $\mathrm{1_{01}}$ line probes the total column, while higher excitation lines probe the higher density gas (e.g., clumps).

}
 
\maketitle

\section{\label{sec:intro}Introduction}
The study of water is of great interest (e.g., Ceccarelli et al. 1996). Not only because $\mathrm{H_2O}$ possesses a very large number of strong far-infrared (FIR) transitions, but also because it plays a crucial role in the chemistry of molecular clouds (Bergin et al. 1998) and can act as an important coolant in dense molecular clouds (Neufeld et al. 1995), in shocks (Draine et al. 1983; Neufeld $\&$ Melnick 1987) and circumstellar outflows (Chen $\&$ Neufeld 1995). Because of its high abundance and easy excitation in warm interstellar and circumstellar environments, water is a powerful tool to probe astrophysical conditions in a broad variety of sources, from protostars to molecular clouds.\\
Nonetheless, the large attenuation by terrestrial $\mathrm{H_2O}$ molecules has made ground-based observations nearly impossible. Therefore, observations from space-born facilities are needed (e.g., Neufeld et al. 1999; Truong-Bach et al. 1999; Wright et al. 2000).\\
The launch of the {\em Infrared Space Observatory} (ISO) has made it possible to look for emission and absorption lines in warm ($\sim$ 100 \mbox{K}) molecular gas near young stellar objects (van Dishoeck et al. 1999), while the {\em Submillimeter Wave Astronomy Satellite} (SWAS) has probed \element[][]{H_2O} and \element[][]{O_2} residing in cold molecular clouds (Melnick et al. 2000), as has Odin (Hjalmarson et al. 2003). The future launch of Herschel, with on board the {\em Heterodyne Instrument for the Far Infrared} (HIFI), will provide even better information on the physical and chemical conditions in molecular clouds by means of observing many water lines with higher angular resolution and sensitivity.\\
In this paper we discuss the excitation of water in PDRs. As a testcase we take \object{S140} because it has been probed extensively in the past with SWAS and will be examined in the future with HIFI. The results presented here are obtained with the use of a newly developed escape probability method described in Sect. \ref{sec:escprob} combined with the inhomogeneous numerical code of Spaans (1996), expanded in Spaans $\&$ van Dishoeck (1997). Section \ref{sec:S140} describes briefly \object{S140}. Section \ref{sec:models} is used to describe the results from the 3D- (in)homogeneous models. We summarize and discuss the paper in Sect. \ref{sec:sumanddisc}.

\section{\label{sec:escprob}The escape probability method}
We calculate the transfer of line radiation of ortho- and para-$\mathrm{H_{2}O}$  in a (in)homogeneous 3-dimensional spherical cloud by use of an escape probability approximation. Let ${n_{i}}$$(x,y,z)$ \mbox{$\mathrm{cm}^{-3}$} be the population density of the $i^\mathrm{th}$ level in point $(x,y,z)$. The equations of statistical equilibrium can then be written as:
\begin{eqnarray}
{n_{i}}(x,y,z) {\sum_{j\not= i}^l} {R_{ij}}(x,y,z) & = & {\sum_{j\not= i}^l} {n_{j}}(x,y,z){R_{ji}}(x,y,z),
\label{eq:stateq}
\end{eqnarray}
where $l$ is the total number of levels included. ${R_{ij}}$$(x,y,z)$ is expressible in terms of the Einstein ${A_{ij}}$ and ${B_{ij}}$ coefficients, and  the collisional excitation ($i$ $<$ $j$) and de-excitation rates ($i$ $>$ $j$)  ${C_{ij}}$:
\begin{eqnarray}
{R_{ij}}(x,y,z) = 
\begin{cases}
{A_{ij}}\ + {B_{ij}}<{J_{ij}}>\ +\ {C_{ij}}(x,y,z), & \text{(i $>$ j)}\\
{B_{ij}}<{J_{ij}}>\  +\ {C_{ij}}(x,y,z), & \text{(i $<$ j)}
\end{cases}
\end{eqnarray}
$<$${J_{ij}}(x,y,z)$$>$ is the mean integrated radiation field at frequency ${\nu_{ij}}$ corresponding to the transition from level $i$ to $j$ at position $(x,y,z)$ in the cloud, given by:
\begin{eqnarray}
 <{J_{ij}}(x,y,z)> & = & [1\ -\ \beta_{ij}(x,y,z)]S_{ij}(x,y,z) \nonumber\\
                   &   & + \beta_{ij}(x,y,z){B}_{ij}(\nu_{ij},T_\mathrm{BB})\,,
\label{eq:Jij}
\end{eqnarray}
where ${\beta_{ij}}$(x,y,z) is the probability for a photon to escape and ${S_{ij}}$(x,y,z) is the source function, to be addressed later.
The background radiation ${{B}_{ij}}$(${\nu_{ij}}$,${T_\mathrm{BB}}$) consists of two terms: the 2.7 K microwave background and the infrared emission of dust at a temperature ${T_\mathrm{d}}$ and with an emission optical depth ${\tau_\mathrm{d}(\nu_{ij})}$:
\begin{eqnarray}
{B}_{ij}(\nu_{ij},T_\mathrm{BB}) & = & B(\nu_{ij}, T = 2.7\ \mbox{K}) \nonumber\\
                                 &   & + \tau_\mathrm{d}(\nu_{ij})B(\nu_{ij}, T_\mathrm{d})\,,
\label{eq:Bij}
\end{eqnarray}
where ${\tau_\mathrm{d}(\nu_{ij})}$ = ${\tau_{100 \mu \mathrm{m}}}$(100 ${\mu}$m/${\lambda}$). We adopt a value for ${\tau_{100 \mu \mathrm{m}}}$ of 0.001 (Hollenbach et al. 1991).
If all the photons escape (e.g., ${\beta_{ij}}$ = 1), then  $<$${J_{ij}}(x,y,z)$$>$ is  ${{B}_{ij}}$(${\nu_{ij}}$,${T_\mathrm{BB}}$); if none escape (e.g., ${\beta_{ij}}$ = 0) $<$${J_{ij}}(x,y,z)$$>$ is the local source function ${S_{ij}}$$(x,y,z)$.\\
The source function reads as
\begin{eqnarray}
{S_{ij}}(x,y,z)\ =\ {2h\nu_{ij}^3 \over c^2} {\left[{n_j(x,y,z)g_i \over n_i(x,y,z)g_j} -1 \right]}^{-1}\,,
\label{eq:Sij}
\end{eqnarray}
where ${g_i}$ and ${g_j}$ are the statistical weights of levels $i$ and $j$, respectively.\\
Since the set of $l$ statistical equilibrium equations is not independent, one equation has to be replaced by the conservation equation
\begin{eqnarray}
{n_x} = {\sum_{j = 0}^l} {n_j}\,,
\label{eq:Conserveq}
\end{eqnarray}
where ${n_x}$ is the number density of species $x$ in all levels.\\
In a sphere, the probability that a photon emitted in the transition from level $i$ to level $j$ at position $(x,y,z)$ along a direction $\vec{k}$ escapes the cloud is given by 
\begin{eqnarray}
{\beta_{ij}}(x,y,z,\vec{k}) = {1\ -\ \exp(-\tau_{ij}(x,y,z,\vec{k}))\over \tau_{ij}(x,y,z,\vec{k})}\,,
\label{eq:betaijsphere}
\end{eqnarray}
with ${\tau_{ij}}$ the optical depth in the line. 
Note that the probability for a photon to escape in a point, p$(x,y,z)$, equals the sum of the escape probabilities over all directions, e.g., $\beta_{ij}$$(x,y,z)$ =  $\Sigma_{\vec{k}}$ ${\beta_{ij}}$($x$,$y$,$z$,$\vec{k}$).
The number of directions is arbitrary, but a 6-ray approximation is implemented for the 3D models. Note that for a photon emitted in a plane-parallel medium, the probability to escape is given by the expression
\begin{eqnarray}
{\beta_{ij}}(x,y,z,\vec{k}) = {1\ -\ \exp(-3\tau_{ij}(x,y,z,\vec{k}))\over 3\tau_{ij}(x,y,z,\vec{k})}
\label{eq:betaijplane}
\end{eqnarray}
The optical depth averaged over the line, in direction $\vec{k}$ over a distance s = ${s_2}$(${x_2}$,${y_2}$,${z_2}$) - ${s_1}$(${x_1}$,${y_1}$,${z_1}$) is given by
\begin{eqnarray}
{\tau_{ij}}(x,y,z,\vec{k}) = {A_{ij}c^3\over 8\pi \nu_{ij}^3} \int\limits_{s=s_1}^{s=s_2}{n_i\over \Delta \mathrm{v_d}}\left[{n_jg_i\over n_ig_j} - 1\right]ds,
\label{eq:tauij}
\end{eqnarray}
with $\Delta$${\mathrm{v}_\mathrm{d}}$ = ${(\mathrm{v}_\mathrm{th}^2 + \mathrm{v}_\mathrm{turb}^2)^{1/2}}$ the velocity dispersion. Turbulence in PDRs can be combined with the thermal speed of the gas into a Gaussian distribution. The resultant Doppler profile is
\begin{eqnarray}
{\varphi(x)} = {e^{-x^2}\over \pi^{1/2}}\,,
\label{eq:Dopprof}
\end{eqnarray}
where x = ($\mathrm{\nu}$ - ${\nu_{ij}}$)/${\Delta \nu_\mathrm{d}}$. The Doppler frequency width ${\Delta \nu_\mathrm{d}}$ is given by ${\Delta\nu_\mathrm{d}}$ = ${\Delta \mathrm{v}_\mathrm{d}}$${\nu_{ij} \over \mathrm{c}}$. 
The turbulent dispersion, ${\mathrm{v}_\mathrm{turb}}$, is larger than the thermal velocity of the gas, ${\mathrm{v}_\mathrm{th}}$, and dominates the line broadening in PDRs.\\
Because interpretation of observed water lines requires knowledge of collisional excitation rates responsible for line formation, it is necessary to work with the most accurate collisional rate data available. The collisional rates are taken from Green et al. (1993) for collisions involving up to 8 levels for inelastic collisions between \element[][]{H_2O} and \element[][]{He}. To account for the different reduced mass of the collisional partner (e.g., \element[][]{H_2}), the rate coefficients are scaled by a factor of 1.348. Rate coefficients of para- and ortho \element[][]{H_2}O owing to collisions with ortho- and para-\element[][]{H_2} at kinetic temperatures from 20 to 140 \mbox{K} are taken from Phillips et al. (1996). In addition, rate coefficients for \element[][]{H_2O} in collisions with p-\element[][]{H_2} and o-\element[][]{H_2} for the low temperature regime (5-20 K) are taken from Grosjean et al. (2003) and Dubernet $\&$ Grosjean (2002), respectively. Clumps and their edges dominate the total water emissivity a few magnitudes of extinction into the PDR. The ambient electron abundance in these regions does not exceed 3 $\times$ $10^{-7}$--$10^{-6}$ in our models. Using the \element[][]{H_2}O-electron collision rates presented by Faure et al. (2004) we find that the contribution by electrons to the total collisional excitation rate is less than 1 $\%$ and is therefore ignored in our calculations. Note that when the electron abundance exceeds $10^{-5}$, this contribution is equal to the collisional rates of \element[][]{H_2}O with para-\element[][]{H_2}, and consequently has to be taken into account.\\
In PDRs, the ortho-to-para ratios of \element[][]{H_2} have typical values in the range 1.5--2.2 (Hasegawa et al. 1987; Ramsay et al. 1993; Chrysostomou et al. 1993; Hora $\&$ Latter 1996; Shupe et al. 1998), 
and are not constant because of gas temperature variations within the PDR. Therefore we adopt the expression for the ortho-to-para ratio (OPR), in thermal equilibrium, defined by
\begin{eqnarray}
\mathrm{OPR}= {{(2I_\mathrm{o} + 1)\sum(2J + 1)\exp\left(-{E_\mathrm{o}(J,K_\mathrm{a},K_\mathrm{c})\over kT}\right)}\over{(2I_\mathrm{p} + 1)\sum(2J + 1)\exp\left(-{E_\mathrm{p}(J,K_\mathrm{a},K_\mathrm{c})\over kT}\right)}}\,, 
\label{eq:OPR}
\end{eqnarray}
where $I_\mathrm{o}$ and $I_\mathrm{p}$ are the total nuclear spin, according to the hydrogen nuclear spins being parallel ($I_\mathrm{o}$ = 1, $\uparrow$$\uparrow$) or anti-parallel ($I_\mathrm{p}$ = 0, $\uparrow$$\downarrow$). The sum in the numerator (denominator) extends over all ortho (para) levels $({J},K_\mathrm{a},K_\mathrm{c})$ (Mumma et al. 1987). A similar expression holds for the ortho- and para-\element[][]{H_2O} OPR, for which the statistical equilibrium (OPR = 3) value is attained at temperatures exceeding $\sim$ 60 K, but will differ from the statistical equilibrium value for lower temperatures.\\
This escape probability method is the first step towards a fully 3D-Monte Carlo radiative transfer code that calculates line intensities, as well as line profiles in different astrophysical environments.

\section{\label{sec:S140}The molecular cloud \object{S140}}
The molecular cloud lies at a distance of $\sim$ 900 \mbox{pc} (Crampton $\&$ Fisher 1974) and is illuminated from the south-western side by the BOV star \object{HD 211880}. The star is located about $\sim$ 7$'$ (1.85 \mbox{pc}) from the edge of the cloud. The cloud extends over more than 30$'$ (8 \mbox{pc}) and contains a dense core in which star formation is occuring. \\
The molecular cloud consists of two parts: (i) the extended molecular cloud (EMC), and (ii) the interface region near the south-west edge of the cloud (PDR). Here focus is on the interface region, as this region is a far more interesting object to probe the behaviour of water with its high \element[][]{H_2O} abundance relative to the extended molecular cloud.\\
In the past, models were neither able to reproduce the observations of extended emission of $[$\ion{C}{i}$]$ (Keene et al. 1985) and  $[$\ion{C}{ii}$]$ (Stutzki et al. 1988) deep into the cloud, nor the intense \element[][13]{CO} 6-5 emission in some regions (Graf et al. 1990). This information, together with the \element[][]{CO} maps by Falgarone $\&$ Perault (1988), led to the suggestion that interstellar clouds have a very inhomogeneous structure, so that the ultraviolet radiation can penetrate much deeper into the cloud.\\
Most previous models of \object{S140} (e.g., Ashby et al. 2000) have assumed a spherically symmetric cloud with temperature and density power laws. Even though they can reproduce well the observed value of the intensity of the 557 GHz ${1_{10}}$ $\rightarrow$ ${1_{01}}$ ground-state transition of ortho-\element[][]{H_2}\element[][16]{O}, being acquainted with the inhomogeneous structure of molecular clouds from observations and in order to explain and predict more accurately various line transitions, we make use of density, temperature and abundance (relative to  \element[][]{H_2}) distributions as calculated with the inhomogeneous code of Spaans (1996). This code is described first in Spaans (1996) and modified in Spaans $\&$ van Dishoeck (1997). This is the first time an attempt is made to model the water emission in \object{S140} in an inhomogeneous way.\\
The Monte Carlo code calculates the chemical structure and thermal balance in PDRs, as well as the distribution of the water abundance and molecular hydrogen density simultaneously. A chemical network including 291 reactions between 51 species consisting of the elements hydrogen, carbon (\element[][12]{C}), oxygen, iron, magnesium, and PAHs (with charges -1, 0, +1) is used. The gas abundances for \object{S140}, are taken to be 2.0 $\times$ $10^{-4}$, 5.0 $\times$ $10^{-4}$, 2.5 $\times$ $10^{-7}$ and 1.3 $\times$ $10^{-6}$ for C, O, Fe and Mg, respectively. For carbon this implies that about 50$\%$ of the solar abundance is present in the gas phase, which is comparable with the values found in diffuse clouds (Cardelli et al. 1993). Mechanisms for converting the ultraviolet flux into gas heating are (1) photoelectric emission from small dust grains and large molecules like Polycyclic Aromatic Hydrocarbons (PAHs) (Bakes $\&$ Tielens 1994; Spaans et al. 1994), (2) neutral carbon ionization, (3) \element[][]{H_2} photodissociation, (4) FUV pumping followed by collisional de-excitation of vibrationally excited \element[][]{H_2} in the two level approximation, and (5) \element[][]{H_2} formation from FUV-produced hydrogen atoms on grain surfaces. In addition, (6) cosmic-ray heating may also contribute deep into the cloud, and (7) \ion{O}{i} may heat the gas through the collisional de-excitation of the $\mathrm{^3P_1}$ fine structure level excited by (dust) continuum radiation.\\
For the cooling, the infrared fine-structure lines from atoms \element[][]{C}, \element[+]{C}, \element[][]{O}, \element[][]{Si}, and \element[][]{Fe}, if present in the chemical network, have been taken into account. The lowest 20 rotational levels of \element[][]{CO} are calculated in statistical equilibrium and their contribution to the cooling is included. \element[][]{H_2} rotational cooling is considered as well as cooling due to water through a cooling function (Neufeld $\&$ Melnick 1987). The contribution of water to the total cooling rate amounts to 10--20$\%$ around the clump density of $10^5$ \mbox{$\mathrm{cm}^{-3}$} and for temperatures of 40--60 K. The CO molecule dominates the cooling for these conditions.\\
The computed temperature, density and water abundance distribution is then used as input for the water line transfer code described previously in this paper. This approach leads to a more accurate treatment of the complex chemical and thermal structure of the PDR, than when a temperature and \element[][]{H_2O} abundance are simply imposed.
 
\section{\label{sec:models}Models}
This section consists of two parts. The first part describes the results from the homogeneous model, the second part describes the inhomogeneous models and the results thereof.
\subsection{Homogeneous 3D model}

\begin{figure}
\includegraphics[width=3.7cm]{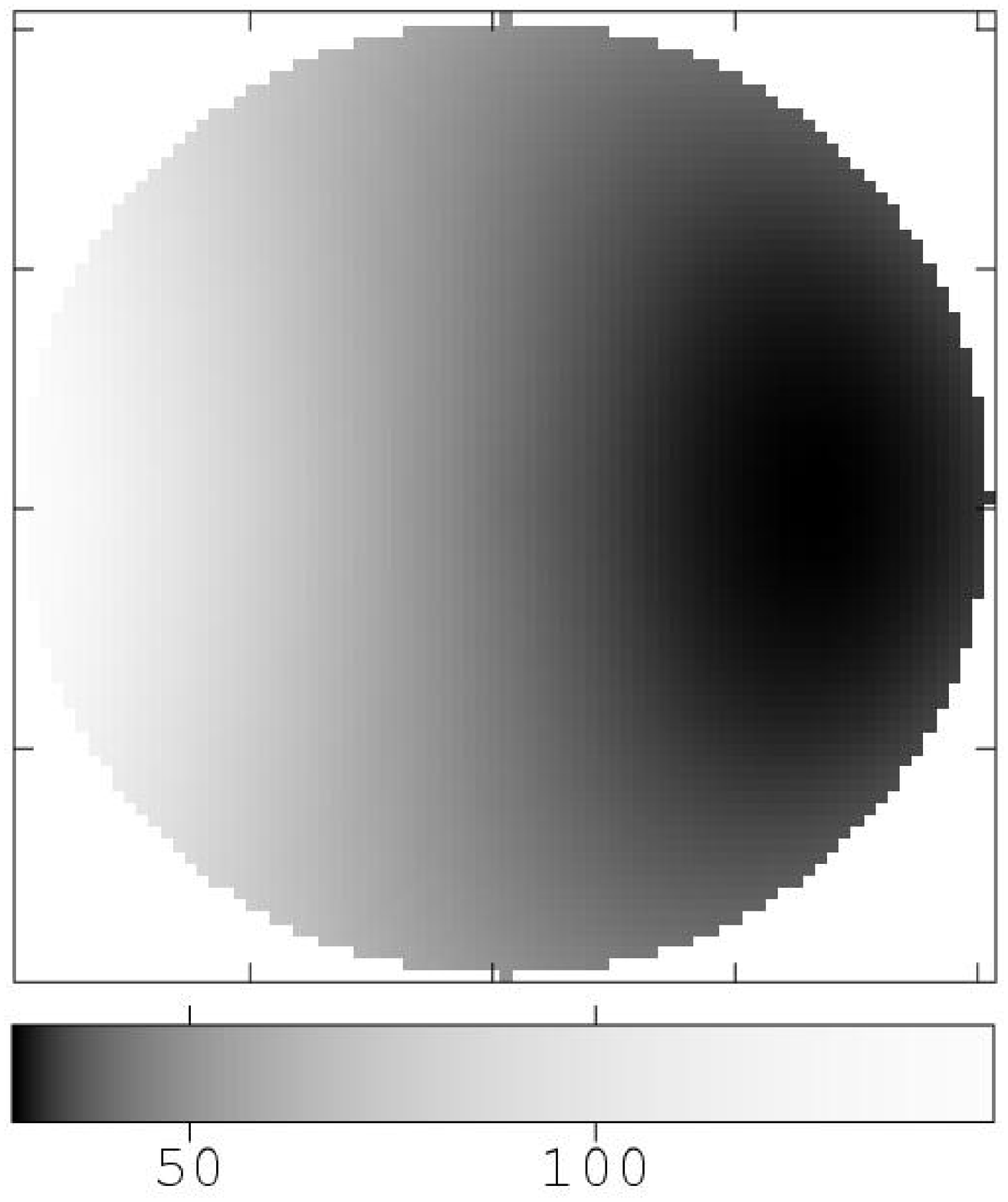}
\includegraphics[width=3.7cm]{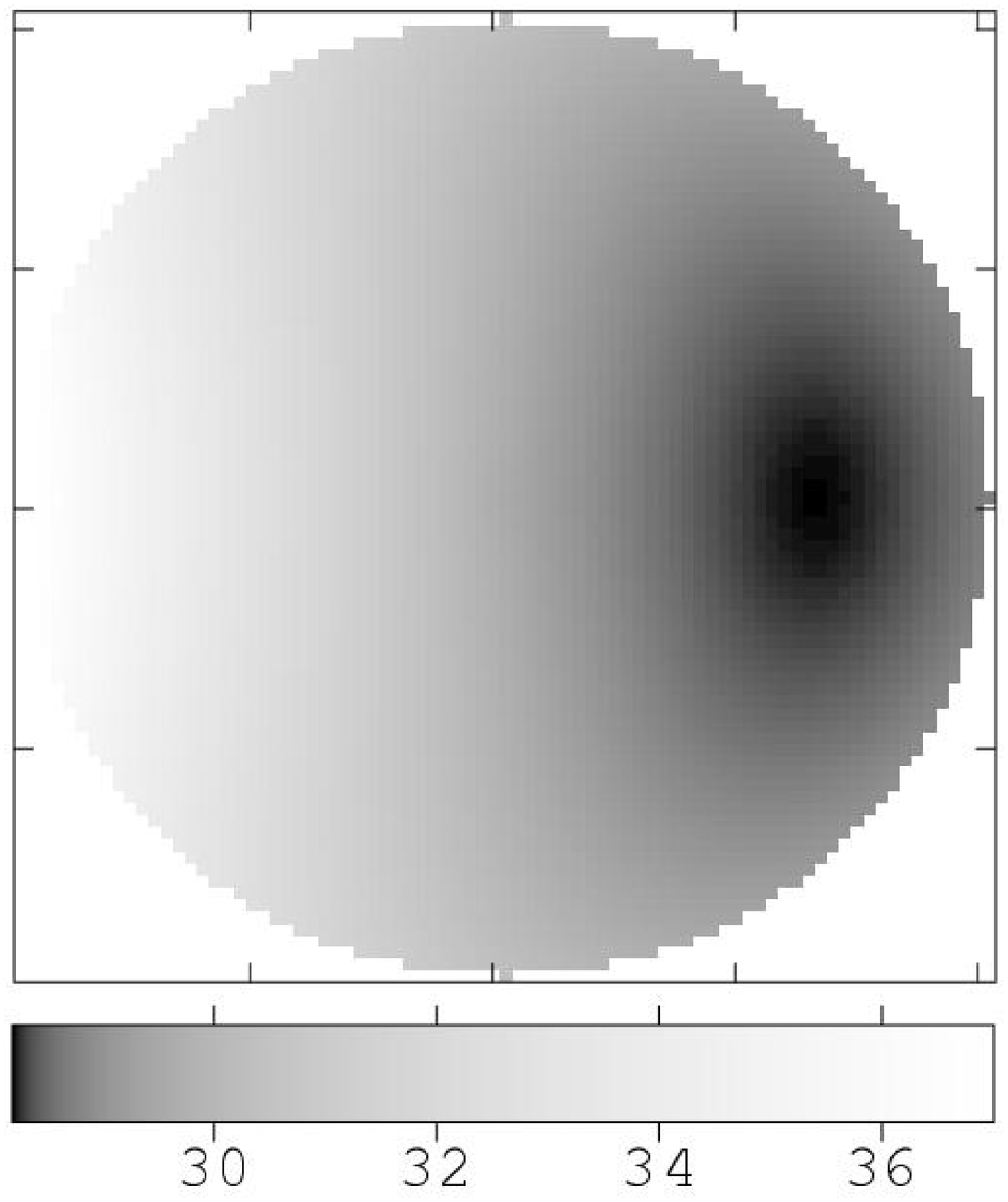}
\caption{Gas (left) and dust (right) temperatures (K) when taking a vertical slice through the centre of the sphere in case of a homogeneous model. The star
HD 21880 is located at the left of both images.}
\label{fig:temp}
\end{figure}

\begin{figure}
\includegraphics[width=3.7cm]{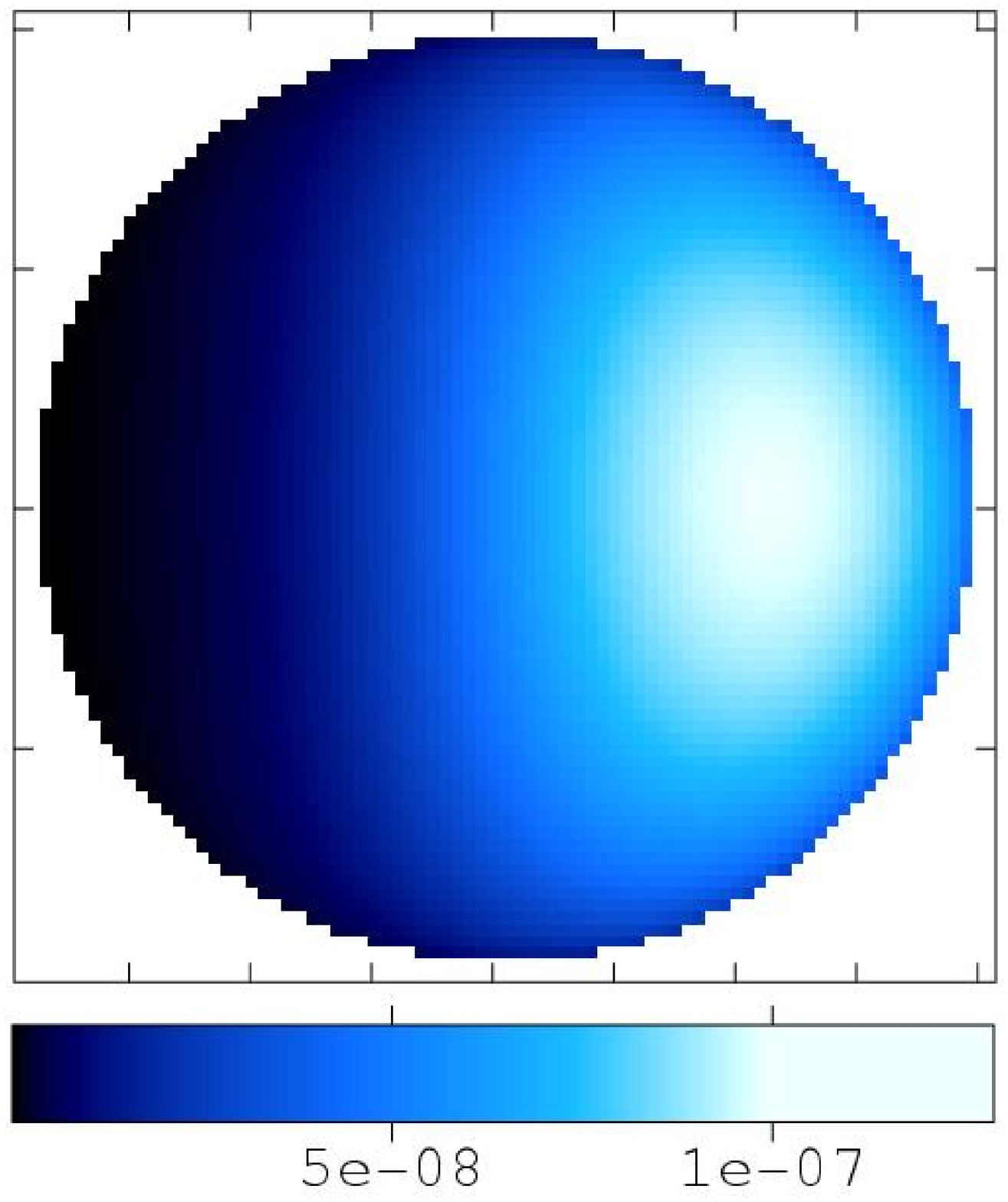}
\includegraphics[width=3.7cm]{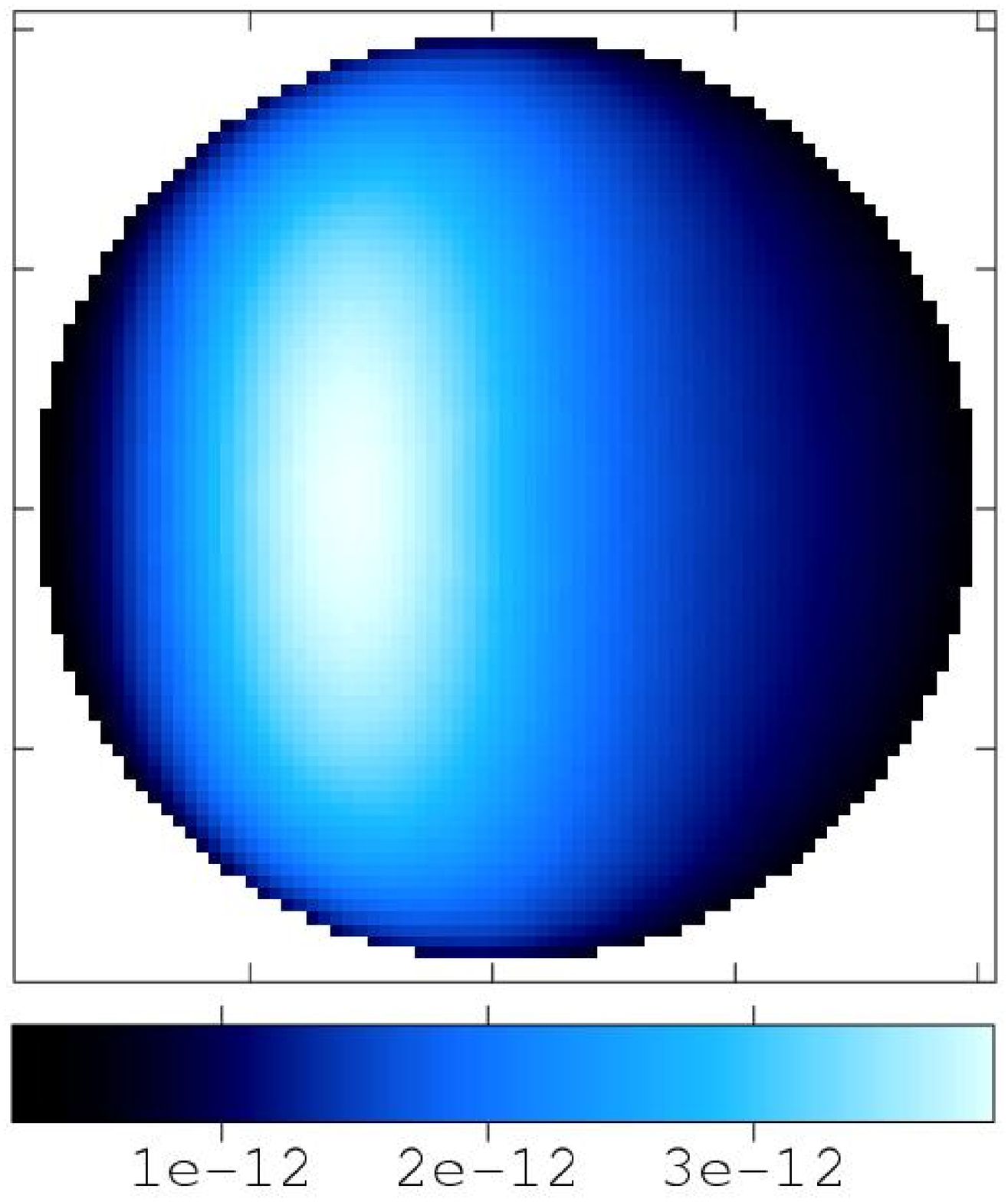}\\\\
\includegraphics[width=3.7cm]{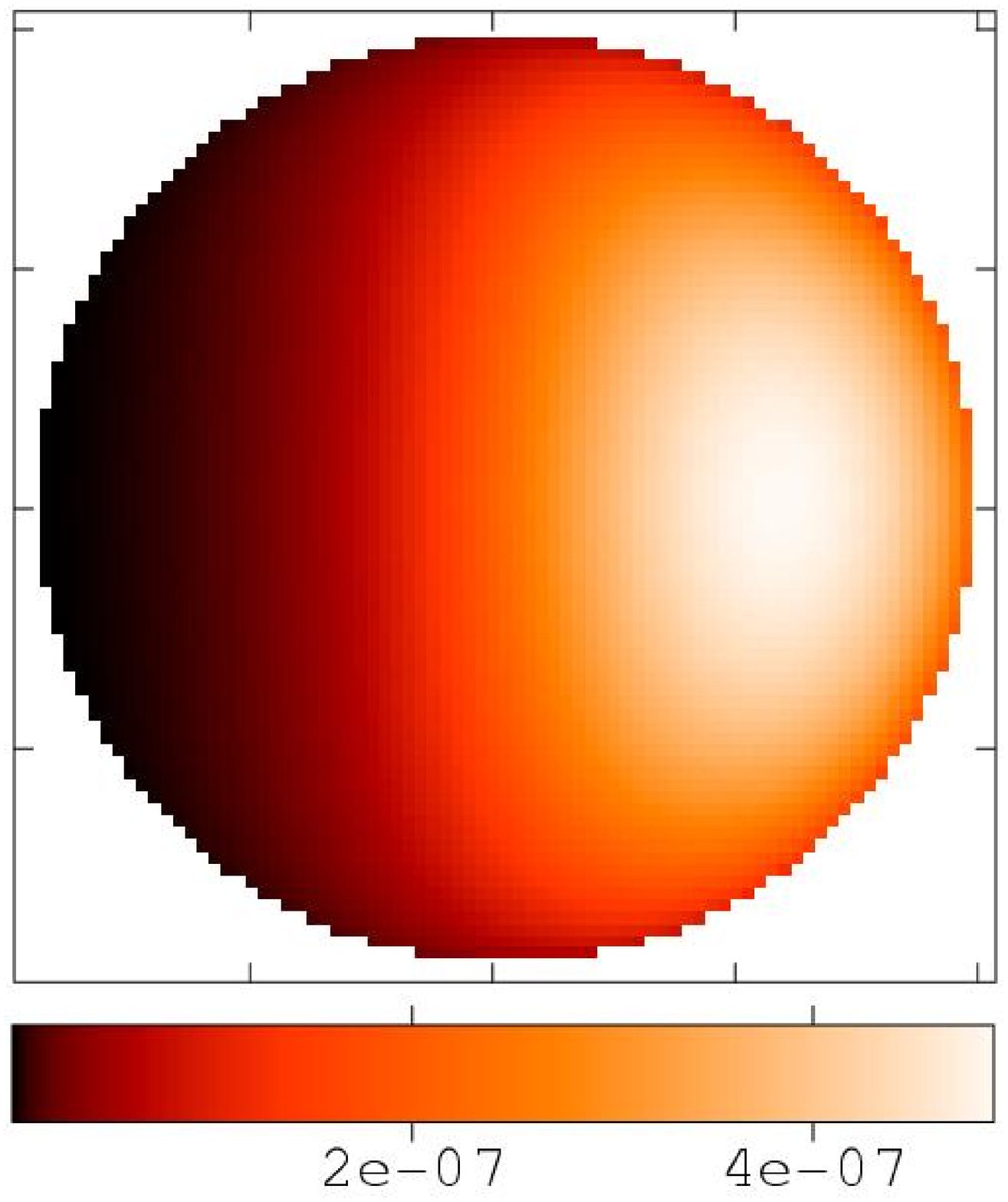}
\includegraphics[width=3.7cm]{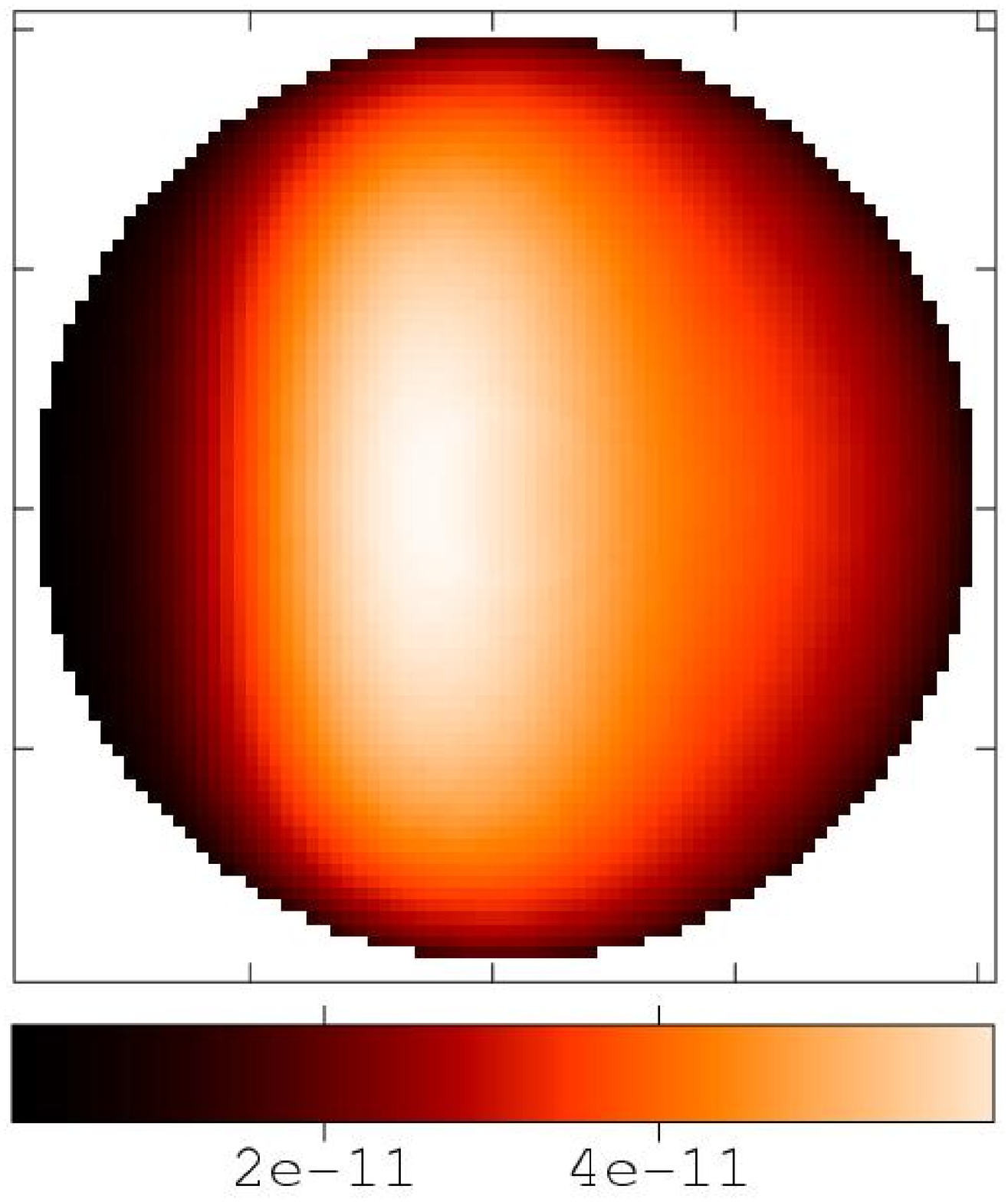}
\caption{Distribution of the intensities (\mbox{$\mathrm{erg}\ \mathrm{cm^{-2}}\ \mathrm{s}^{-1}\ \mathrm{sr^{-1}}$}) for various transitions of ortho-(blue/top) and para-(red/bottom) \element[][]{H_2O}. The top left panel depicts the $1_{10}$ $\rightarrow$ $1_{01}$ transition, the top right the $3_{12}$ $\rightarrow$ $2_{21}$ transition. Bottom left is $1_{11}$ $\rightarrow$ $0_{00}$, and bottom right is $2_{11}$ $\rightarrow$ $2_{02}$. The star HD 211880 is located to the left of the Figures.
\label{fig:2dhomdistr}}
\end{figure}

\begin{table*} 
\begin{minipage}[b]{\columnwidth}
\caption[]{Homogeneous model}
\label{tab:3dhom}
\centering
\renewcommand{\footnoterule}{}
\begin{tabular}{cccccccc}
\hline
${\lambda_\mathrm{ul}}$ & transition   &${E_\mathrm{ul}}$ & ${A_\mathrm{ul}}$ &average\footnote{Is the average intensity over the whole cloud.} & peak & average  & ${n_\mathrm{cr}}$\footnote{Critical density evaluated at 100 \mbox{K}.} \\
 & & & & intensity& intensity& around&\\
 &  &  & & & &peak intensity\footnote{Is the intensity averaged over the HIFI FWHM beam size ({\em http://www.sron.rug.nl/hifiscience/}) for a beam centered on the peak emission.}  &\\
$[$$\mu$m$]$ &  &$[$K$]$&$[$s$]$ & $[$\mbox{$\mathrm{erg}\ \mathrm{cm}^{-2}\ \mathrm{s}^{-1}\ \mathrm{sr}^{-1}$}$]$ &$[$\mbox{$\mathrm{erg}\ \mathrm{cm}^{-2}\ \mathrm{s}^{-1}\ \mathrm{sr}^{-1}$}$]$ & $[$\mbox{$\mathrm{erg}\ \mathrm{cm}^{-2}\ \mathrm{s}^{-1}\ \mathrm{sr}^{-1}$}$]$ &$[$\mbox{$\mathrm{cm}^{-3}$}$]$\\
\hline\hline
538.2 &o-\element[][]{H_2O} $1_{10}$ $\rightarrow$ $1_{01}$ & 26.7  & 3.458 $\times$ $10^{-3}$& 4.00 $\times$ $10^{-8}$ & 1.00 $\times$ $10^{-7}$ & 9.20 $\times$ $10^{-8}$  & 3.5 $\times$ $10^7$\\
179.5 &o-\element[][]{H_2O} $2_{12}$ $\rightarrow$ $1_{01}$ & 80.2  & 5.593 $\times$ $10^{-2}$& 5.66 $\times$ $10^{-7}$ & 1.41 $\times$ $10^{-6}$ & 1.37 $\times$ $10^{-7}$  & 5.0 $\times$ $10^8$\\
108.1 &o-\element[][]{H_2O} $2_{21}$ $\rightarrow$ $1_{10}$ & 133.2 & 2.564 $\times$ $10^{-1}$& 8.26 $\times$ $10^{-10}$ & 2.00 $\times$ $10^{-9}$ & 1.88 $\times$ $10^{-9}$ & 3.6 $\times$ $10^{9}$\\
180.5 &o-\element[][]{H_2O} $2_{21}$ $\rightarrow$ $2_{12}$ & 79.7  & 3.058 $\times$ $10^{-2}$& 5.90 $\times$ $10^{-11}$ & 1.43 $\times$ $10^{-10}$& 1.35 $\times$ $10^{-10}$  & 6.6 $\times$ $10^{8}$\\
174.6 &o-\element[][]{H_2O} $3_{03}$ $\rightarrow$ $2_{12}$ & 82.4  & 5.048 $\times$ $10^{-2}$& 5.70 $\times$ $10^{-10}$ & 1.04 $\times$ $10^{-9}$ & 8.98 $\times$ $10^{-10}$  & 7.7 $\times$ $10^8$\\
260.0 &o-\element[][]{H_2O} $3_{12}$ $\rightarrow$ $2_{21}$ & 55.4  & 2.634 $\times$ $10^{-3}$& 1.86 $\times$ $10^{-12}$ & 4.04 $\times$ $10^{-12}$ & 3.78 $\times$ $10^{-12}$  & 4.8 $\times$ $10^7$\\
273.2 &o-\element[][]{H_2O} $3_{12}$ $\rightarrow$ $3_{03}$ & 52.7  & 1.648 $\times$ $10^{-2}$& 1.11 $\times$ $10^{-11}$ & 2.40 $\times$ $10^{-11}$ & 2.26 $\times$ $10^{-11}$ & 2.8 $\times$ $10^8$\\
 & & & &  &\\
\hline
\hline
 & & & &  &\\ 
269.3 &p-\element[][]{H_2O} $1_{11}$ $\rightarrow$ $0_{00}$ & 53.5  & 1.842 $\times$ $10^{-2}$& 1.76 $\times$ $10^{-7}$ & 4.94 $\times$ $10^{-7}$ & 4.54 $\times$ $10^{-7}$  & 1.6 $\times$ $10^8$\\
303.4 &p-\element[][]{H_2O} $2_{02}$ $\rightarrow$ $1_{11}$ & 47.5  & 5.835 $\times$ $10^{-2}$& 1.78 $\times$ $10^{-9}$ & 4.42 $\times$ $10^{-9}$ & 4.18 $\times$ $10^{-9}$  & 8.9 $\times$ $10^{7}$\\
101.0 &p-\element[][]{H_2O} $2_{20}$ $\rightarrow$ $1_{11}$ & 142.6 & 2.607 $\times$ $10^{-1}$& 1.93 $\times$ $10^{-10}$ & 5.40 $\times$ $10^{-10}$ & 5.00 $\times$ $10^{-10}$  & 3.8 $\times$ $10^{9}$\\
398.6 &p-\element[][]{H_2O} $2_{11}$ $\rightarrow$ $2_{02}$ & 36.1  & 7.062 $\times$ $10^{-3}$& 3.34 $\times$ $10^{-11}$ & 6.40 $\times$ $10^{-11}$ & 5.64 $\times$ $10^{-11}$ & 8.7 $\times$ $10^{7}$\\
\hline
\end{tabular}
\end{minipage}
\end{table*}

The incident radiation field, taken to be ${I_\mathrm{UV}}$ = 140 with respect to the Draine (1978) field, is consistent with the enhancement at the edge of the cloud near the BOV star HD 211880. An isotropic component, ${I_\mathrm{UV}}$ = 1, of the average interstellar radiation field has also been taken into account. For the homogeneous PDR model we adopt a generic mean total hydrogen density of $n_\mathrm{H}$ = $n$(\element[][]{H}) + 2$n$(\element[][]{H_2}) = 2 $\times $ $10^4$ \mbox{$\mathrm{cm}^{-3}$} over a region 0.5 \mbox{pc} in extent, or $A_\mathrm{V}$ $\approx$ 20 mag, in agreement within a factor $\sim$ 1.5--2 with observations (Hayashi et al. 1985; Zhou et al. 1994). For areas where \element[][]{H_2O} is present, hydrogen is totally molecular (the main part of the UV radiation is already absorbed), and therefore $n_\mathrm{H}$ = 2n(\element[][]{H_2}). The grid size is taken to be 81 $\times$ 81 $\times$ 81, corresponding to a resolution of 0.006 \mbox{pc}.\\ 
The temperature distribution for the gas and dust is shown in Fig. \ref{fig:temp}. 
The gas (dust) temperature is $\approx$ 150 \mbox{K} (40 \mbox{K}) at the western edge of the cloud, and drops to about 30 \mbox{K} (30 \mbox{K}) deeper into the cloud. The dust temperature stays far below the gas temperature at the west side of the cloud, but becomes slightly larger or equal to the gas temperature farther into the cloud. It is worth noting that dust plays an important role in the excitation of water. It absorbs the interstellar FUV and visible radiation field and emits the absorbed energy in the far infrared, offering a source of excitation for water. 
Computed water abundances vary from $\sim$ $10^{-10}$ to $\sim$ $10^{-6}$, relative to the hydrogen density $n_\mathrm{H}$, with an average abundance of 1.4 $\times$ ${10^{-7}}$. At the west side of the cloud, where the star is located, the abundance is lowered due to photodissociation of water by the FUV radiation field:
\begin{center}
\element[][]{H_2O} + $\mathrm{h\nu}$ $\rightarrow$ \element[][]{H} + \element[][]{OH},\\
\element[][]{OH} + $\mathrm{h\nu}$ $\rightarrow$ \element[][]{O} + \element[][]{H}
\end{center}
 Deeper into the cloud, the radiation field is attenuated by dust, resulting in a higher abundance.\\
At low temperatures \element[][]{H_2O} is formed mainly through the dissociative recombination of the \element[+]{H_3O} ion:
\begin{center}
\element[+]{H_3O} + \element[-]{e} $\rightarrow$ \element[][]{H_2O} + \element[][]{H}
\end{center}
A value of 0.3 is chosen for the branching ratio leading to \element[][]{H_2O} in the dissociative recombination of \element[+]{H_3O} with a rate coefficient of 3.3 $\times$ $10^{-7}$ $(T/300)^{-0.3}$ \mbox{$\mathrm{cm}^{3}\ \mathrm{s}^{-1}$}. The branching ratio is consistent with the results of Vejby-Christensen et al. (1997) but higher than the value of 0.05 suggested by Williams et al. (1996). At temperatures above $\sim$ 300 \mbox{K}, the formation of water is dominated by the neutral-neutral endoergic reaction sequence:
\begin{center}
 \element[][]{O}+ \element[][]{H_2} $\rightleftharpoons$ \element[][]{OH} + \element[][]{H},\\
\element[][]{OH} + \element[][]{H_2} $\rightleftharpoons$ \element[][]{H_2O} + \element[][]{H}
\end{center}
(Elitzur $\&$ de Jong 1978; Elitzur $\&$ Watson 1978). The activation energy barriers are overcome by the thermal energy of the warm gas.\\
The emergent intensity integrated along a line of sight is calculated in the following way:\\
\begin{eqnarray}
{I} = {(4\pi)^{-1}} {\int_o^z} {\Lambda(z')}{dz'}\,, 
\label{eq:Intensity}
\end{eqnarray}
where
\begin{eqnarray}
{\Lambda(\nu_{ij})} = {n_iA_{ij}h\nu_{ij}\beta(\tau_{ij})}\,,
\label{eq:cooling}
\end{eqnarray} 
with ${n_i}$ the population density in the $i^\mathrm{th}$ level, ${A_{ij}}$ the Einstein A coefficient, h${\nu_{ij}}$ the energy difference between the levels $i$ and $j$, ${\tau_{ij}}$ the optical depth averaged over the line, and ${\beta(\tau_{ij})}$ the escape probability at the optical depth ${\tau_{ij}}$ of the line.\\ 
Average intensities, peak intensities, as well as intensities averaged over the HIFI FWHM beam size at the transition frequency of the particular line for a beam centered on the peak emission are listed in Table \ref{tab:3dhom}.  
Figure \ref{fig:2dhomdistr} shows maps of the predicted distribution of the intensities for various transitions for ortho-, and para-\element[][]{H_2O} as we would see them on the sky, convolved with a beam size of 0.3$''$ ($\sim$ the angular extent that corresponds to one gridpoint).\\ 
To interpret the results one has to take into account the distribution of the gas and dust temperature ($T_\mathrm{g}$ and $T_\mathrm{d}$, respectively), the abundance of water in the cloud, the density of the medium, as well as optical depth effects. Since the overall mean density of hydrogen is 2 $\times$ $10^4$ \mbox{$\mathrm{cm}^{-3}$}, far below the critical density for all the transitions (of the order of $10^8$ \mbox{$\mathrm{cm}^{-3}$} or higher), the lines are subthermally excited. This means that collisional de-excitation is less important than spontaneous emission, which results in a distribution of the level populations deviating from LTE.\\
The emission will originate from a depth where $\tau_{ij}$ $\sim$ 1. The place where $\tau_{ij}$ $\sim$ 1 varies for every transition. Optically thick lines, i.e., $\tau_{ij}$ $\ge$ 1, can be used to probe the physical conditions in the outer ($\tau_{ij}$ $<$ 1) layers of the cloud. Lines which are optically thin can be used to probe the physical conditions throughout the cloud. 
A peak in intensity for a certain transition will occur at places in the cloud where the temperature is high enough to excite the line and will depend on the column of \element[][]{H_2O} along the line of sight. Therefore at the edges of the cloud, where the column is rather low, the intensity drops.\\
The o-\element[][]{H_2O} $1_{10}$ $\rightarrow$ $1_{01}$ transition peaks near the east edge of the cloud where the temperature of the gas drops to $\sim$ 60 \mbox{K}, the perfect regime to excite the line. Transitions with a higher energy difference, e.g., o-\element[][]{H_2O} $3_{12}$ $\rightarrow$ $2_{21}$ , have their peak shifted towards the edge of the cloud where the temperature is higher. Note that HIFI, with its high angular resolution will be able to see this shift. Depending on the frequency of the  transition, the beam will be $\sim$ 5 times smaller than the PDR.\\
The o-\element[][]{H_2O} $2_{12}$ $\rightarrow$ $1_{01}$ transition is the strongest we found, with an average intensity of $\approx$ 5 $\times$ \mbox{$10^{-7}\ \mathrm{erg}\ \mathrm{cm}^{-2}\ \mathrm{s}^{-1}\ \mathrm{sr}^{-1}$}.

\subsection{Inhomogeneous 3D model with the EMC}

\begin{figure}
\includegraphics[width=3.7cm]{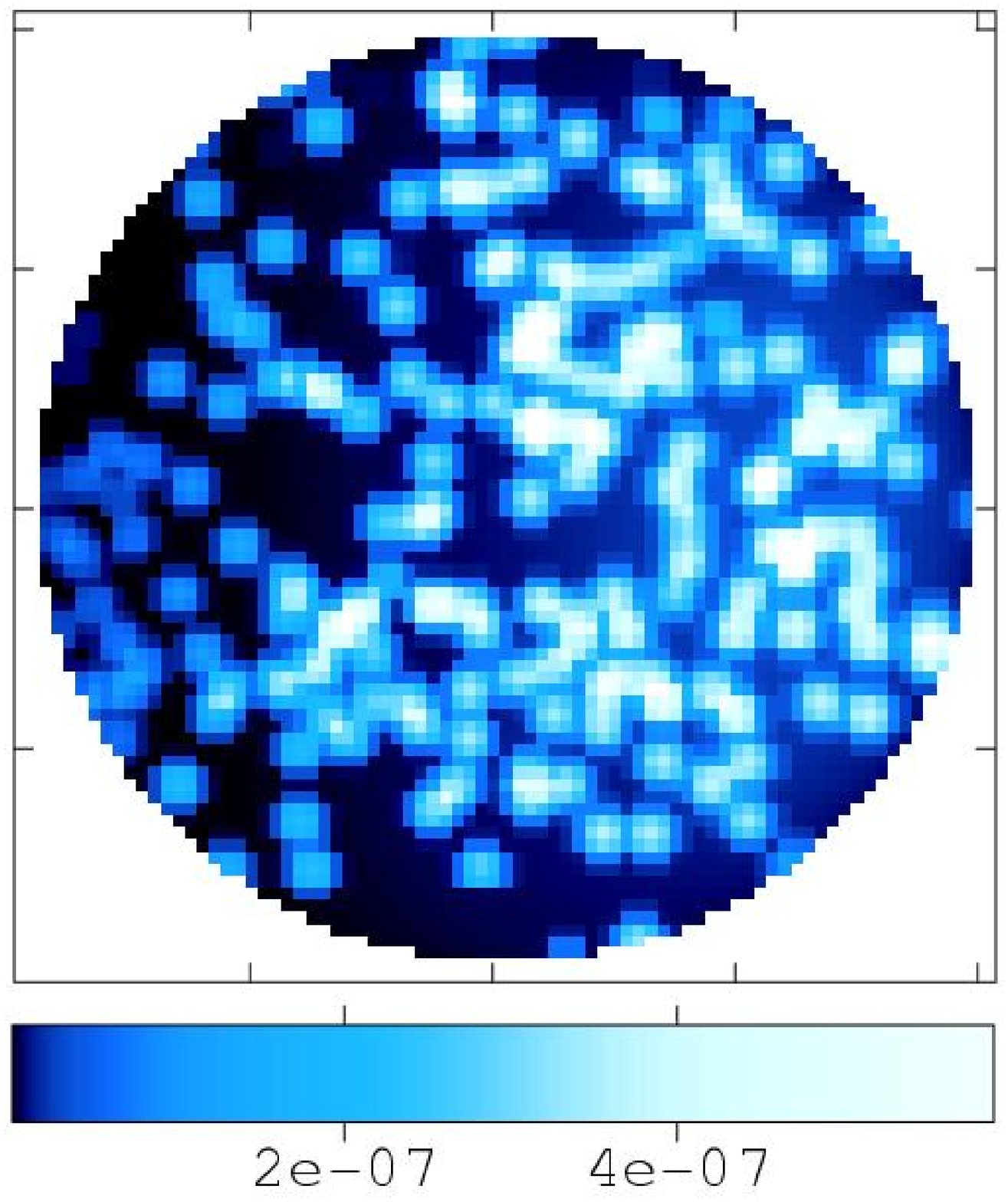}
\includegraphics[width=3.7cm]{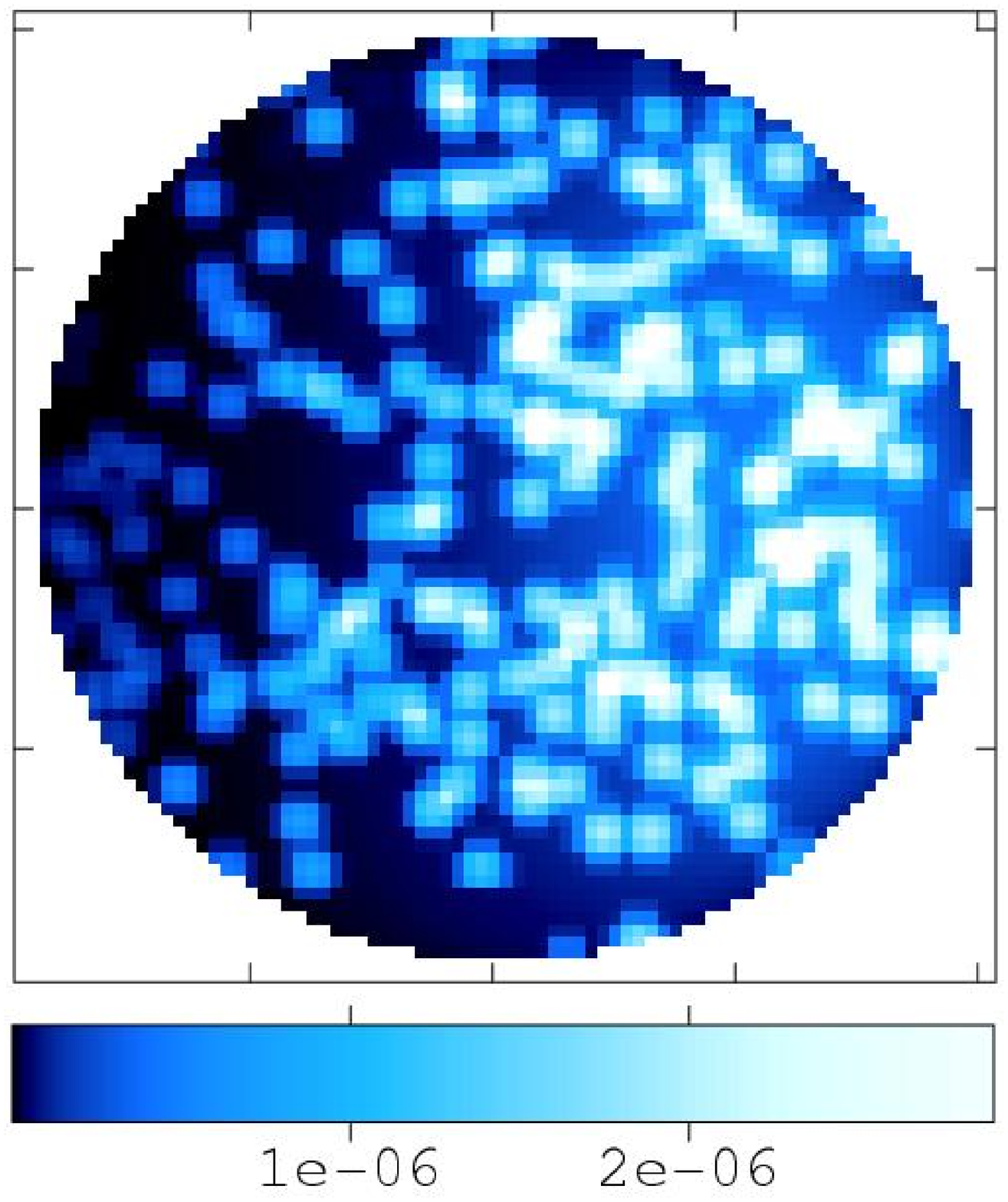}\\\\
\includegraphics[width=3.7cm]{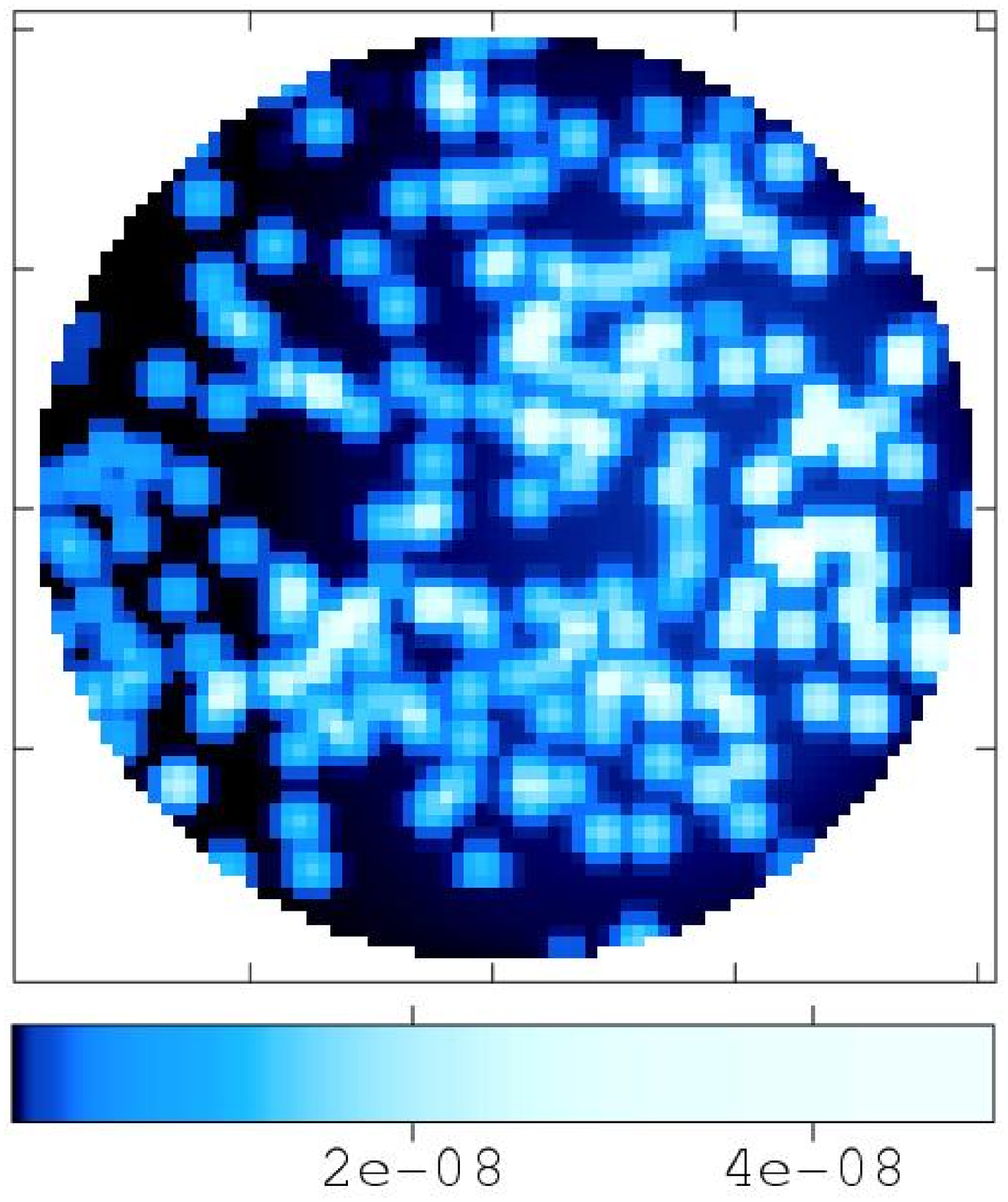}
\includegraphics[width=3.7cm]{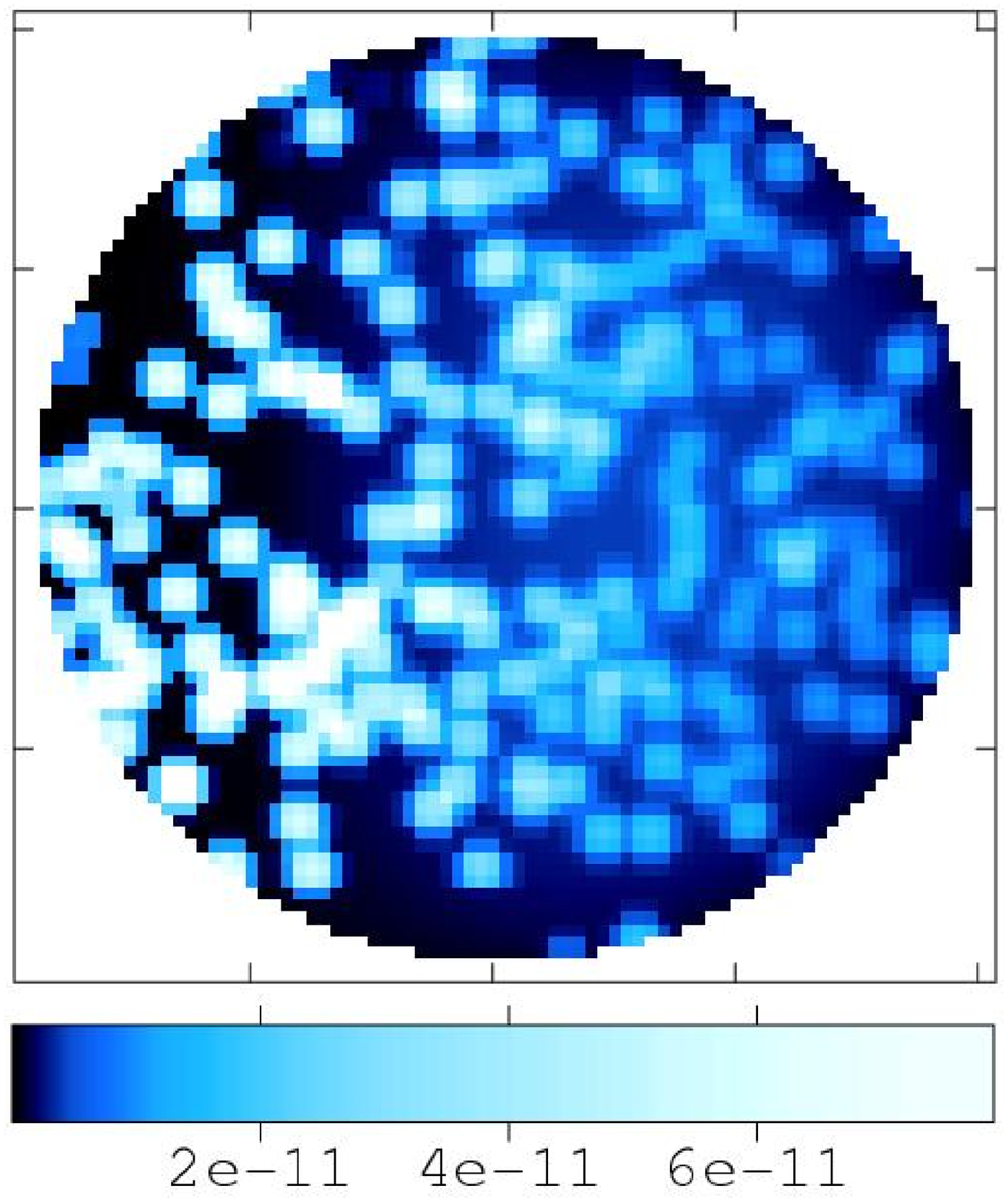}
\caption{Distribution of the  intensities (\mbox{$\mathrm{erg}\ \mathrm{cm^{-2}}\ \mathrm{s}^{-1}\ \mathrm{sr^{-1}}$}) for various transitions of ortho-\element[][]{H_2O} using the PDR I model. The top left panel depicts the $1_{10}$ $\rightarrow$ $1_{01}$ transition, the top right the $2_{12}$ $\rightarrow$ $1_{01}$ transition. Bottom left is $2_{21}$ $\rightarrow$ $1_{10}$, and bottom right is $3_{12}$ $\rightarrow$ $2_{21}$. The star HD 211880 is located to the left of the Figures.\label{fig:3dinhomorthodistr}}
\end{figure}

\begin{figure}
\includegraphics[width=3.7cm]{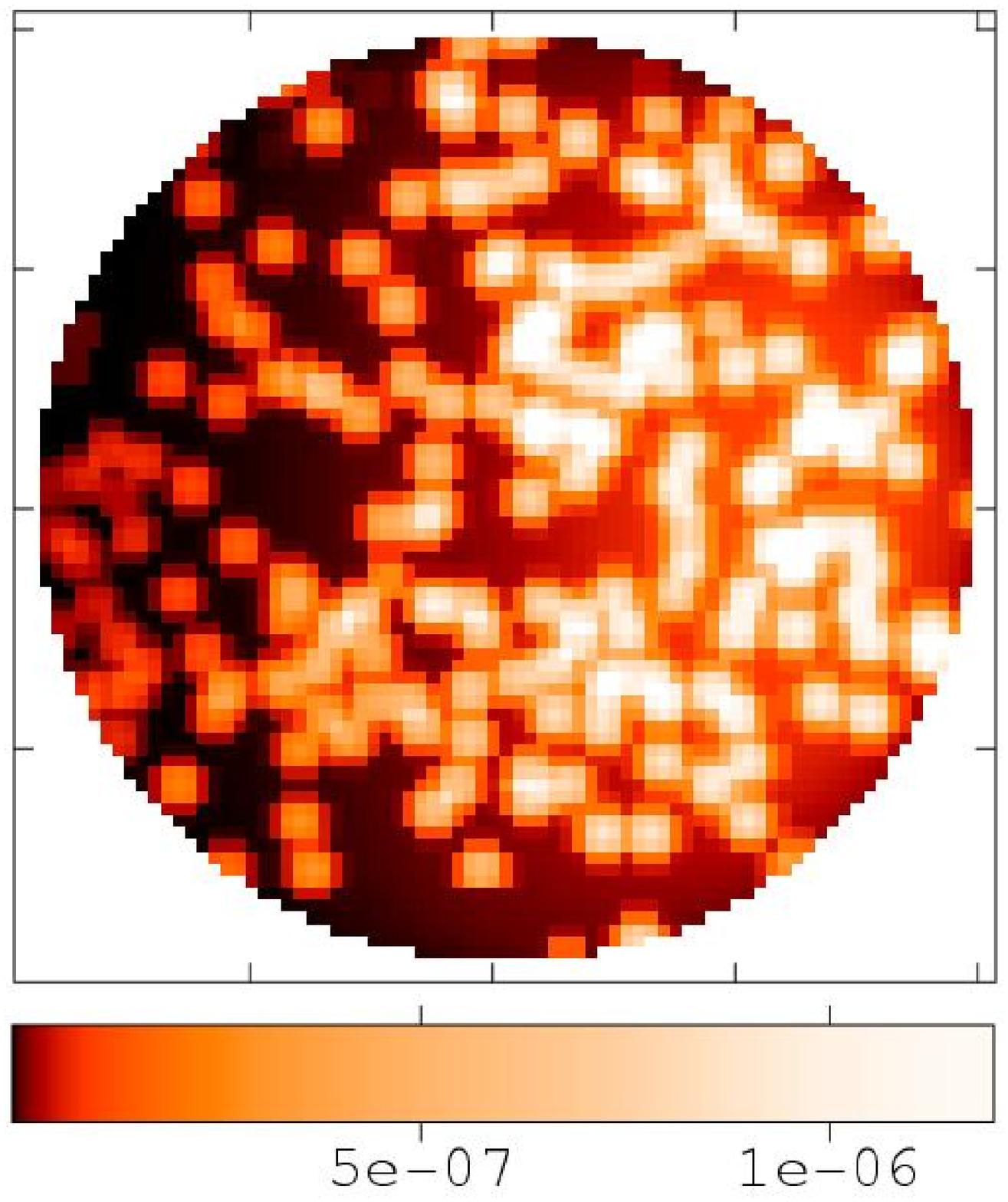}
\includegraphics[width=3.7cm]{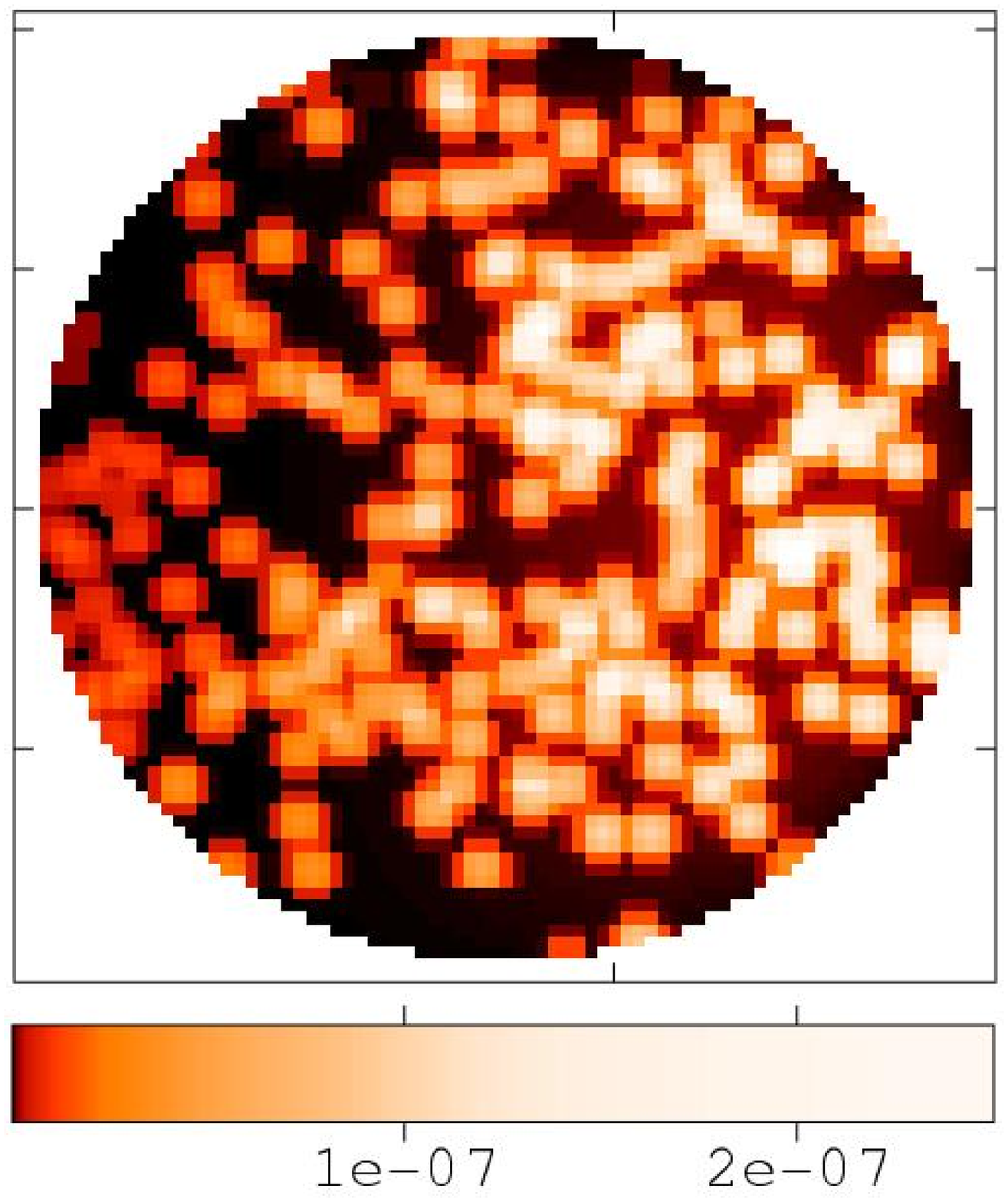}\\\\
\includegraphics[width=3.7cm]{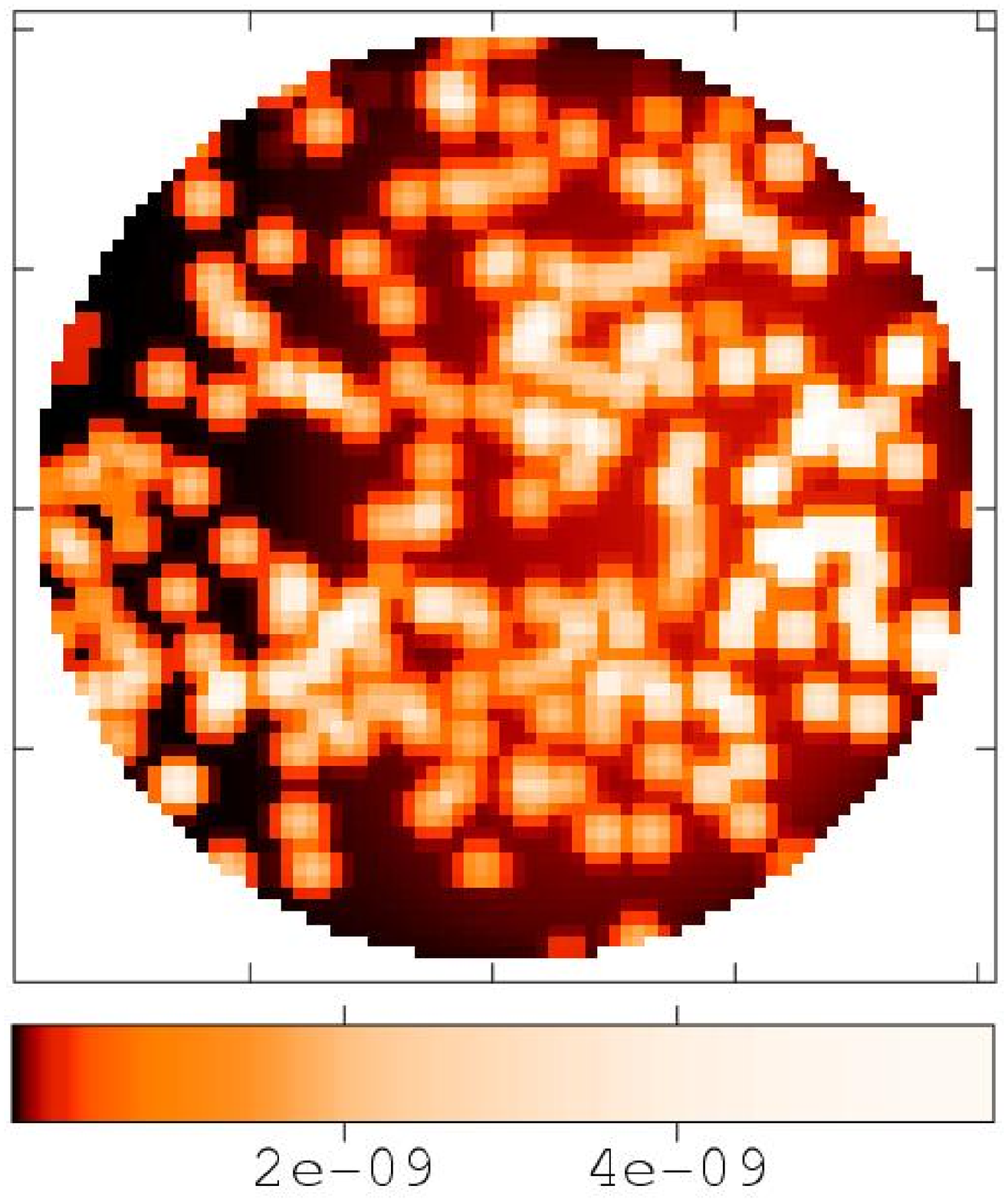}
\includegraphics[width=3.7cm]{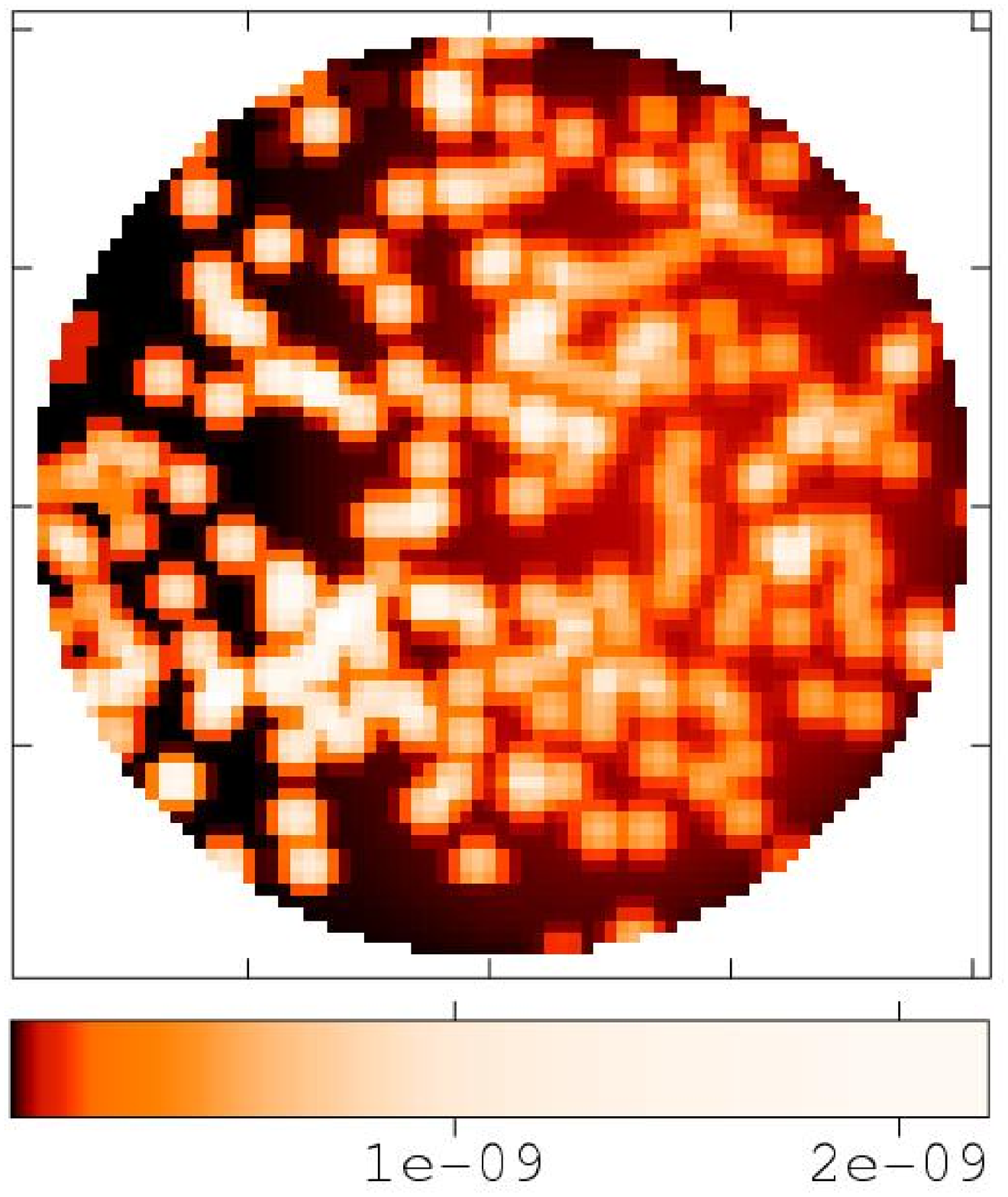}
\caption{Distribution of the intensities (\mbox{$\mathrm{erg}\ \mathrm{cm^{-2}}\ \mathrm{s}^{-1}\ \mathrm{sr^{-1}}$}) for various transitions of para-\element[][]{H_2O} using the PDR I model. The top left panel depicts the $1_{11}$ $\rightarrow$ $0_{00}$ transition, the top right the $2_{02}$ $\rightarrow$ $1_{11}$ transition. Bottom left is $2_{20}$ $\rightarrow$ $1_{11}$, and bottom right is $2_{11}$ $\rightarrow$ $2_{02}$. The star HD 211880 is located to the left of the Figures.\label{fig:3dinhomparadistr}}
\end{figure}

\begin{figure}
\includegraphics[width=7.3cm]{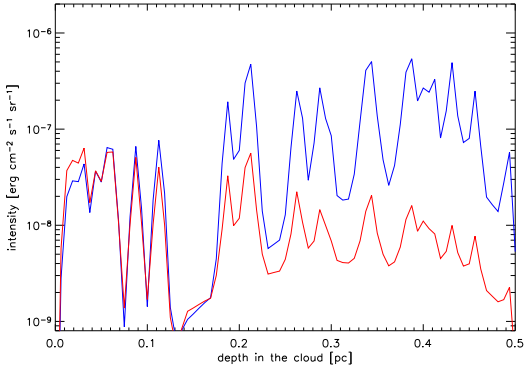}
\caption{ Change in intensity along a W-E cut for the ortho $1_{10}$ $\rightarrow$ $1_{01}$ (blue/top), and ortho $3_{12}$ $\rightarrow$ $2_{21}$ (red/bottom) transition as seen in Fig. \ref{fig:3dinhomorthodistr}. The $3_{12}$ $\rightarrow$ $2_{21}$ transition is multiplied by 1000. The intensity, for the $1_{10}$ $\rightarrow$ $1_{01}$ rises until it reaches its maximum near the east side of the PDR and slightly levels off towards the edge. The $3_{12}$ $\rightarrow$ $2_{21}$ transition reaches its peak near the west side of the PDR and levels off towards the east side.}
\label{fig:WEcutinten}
\end{figure}

\begin{table*}
\begin{minipage}[b]{\columnwidth}
\renewcommand{\footnoterule}{}
\caption{Model parameters}
\label{tab:parameters}
\centering
\begin{tabular}{lcccccccc}
\hline  
Model & size & ${n_\mathrm{c}}$ & $n_\mathrm{ic}$ & $F$ & $R_\mathrm{ic}$& $l_\mathrm{c}$ & $\mathrm{A(\element[][]{H_2O})}$\footnote{Volume averaged water abundance.} & $\Delta$$\mathrm{v}_\mathrm{d}$ \\
 & $[$\mbox{pc}$]$ & $[$\mbox{$\mathrm{cm}^{-3}$}$]$ & $[$\mbox{$\mathrm{cm}^{-3}$}$]$& $[$\% $]$ && $[$\mbox{pc}$]$ & &$[$\mbox{$\mathrm{km}\ \mathrm{s}^{-1}$}$]$\\
\hline\hline
PDR I& 0.5 & 4 $\times$ ${10^5}$ & 1 $\times$ ${10^4}$ & 2.5 & 40& 0.03 & 1.1 $\times$ ${10^{-8}}$ & 1.2\\
PDR II& 0.5 & 2 $\times$ ${10^5}$ & 1 $\times$ ${10^4}$ & 5 & 20& 0.03 & 1.7 $\times$ ${10^{-8}}$ & 1.2\\
PDR III& 0.5 & 6 $\times$ ${10^4}$ & 6 $\times$ ${10^3}$ & 20 &10& 0.03 & 2.1 $\times$ ${10^{-8}}$ & 1.2\\
EMC & 10  &3000  & 300 & 25 & 10 & 0.6 & 6.6 $\times$ ${10^{-10}}$ & 0.7\\
\hline
\end{tabular}
\end{minipage}
\end{table*}

\begin{table*}
\begin{minipage}[b]{\columnwidth}
\caption{Inhomogeneous PDR models}
\label{tab:PDR}
\centering
\renewcommand{\footnoterule}{}
\begin{tabular}{lcccc}
\hline  
Model & transition &average &peak  &average \\
 &  & intensity & intensity& around \\
 &  &  & &peak intensity\footnote{Is the intensity averaged over the HIFI FWHM beam size ({\em http://www.sron.rug.nl/hifiscience/}) for a beam centered on the peak emission.} \\
 & &$[$\mbox{$\mathrm{erg}\ \mathrm{cm}^{-2}\ \mathrm{s}^{-1}\ \mathrm{sr}^{-1}$}$]$ &$[$\mbox{$\mathrm{erg}\ \mathrm{cm}^{-2}\ \mathrm{s}^{-1}\ \mathrm{sr}^{-1}$}$]$ & $[$\mbox{$\mathrm{erg}\ \mathrm{cm}^{-2}\ \mathrm{s}^{-1}\ \mathrm{sr}^{-1}$}$]$ \\
\hline\hline
PDR I &o-\element[][]{H_2O} $1_{10}$ $\rightarrow$ $1_{01}$ & 1.06 $\times$ ${10^{-7}}$  & 1.92 $\times$ ${10^{-6}}$  & 2.20 $\times$ ${10^{-7}}$\\
      &o-\element[][]{H_2O} $2_{12}$ $\rightarrow$ $1_{01}$ & 6.16 $\times$ ${10^{-7}}$  & 6.56 $\times$ ${10^{-6}}$  & 1.39 $\times$ ${10^{-6}}$ \\
      &o-\element[][]{H_2O} $2_{21}$ $\rightarrow$ $1_{10}$ & 5.65 $\times$ ${10^{-9}}$  & 1.71 $\times$ ${10^{-7}}$  & 9.12 $\times$ ${10^{-9}}$\\
      &o-\element[][]{H_2O} $2_{21}$ $\rightarrow$ $2_{12}$ & 4.05 $\times$ ${10^{-10}}$ & 1.23 $\times$ ${10^{-8}}$  & 6.51 $\times$ ${10^{-10}}$\\
      &o-\element[][]{H_2O} $3_{03}$ $\rightarrow$ $2_{12}$ & 4.67 $\times$ ${10^{-9}}$  & 1.20 $\times$ ${10^{-7}}$  & 8.00 $\times$ ${10^{-9}}$\\
      &o-\element[][]{H_2O} $3_{12}$ $\rightarrow$ $2_{21}$ & 1.31 $\times$ ${10^{-11}}$ & 5.38 $\times$ ${10^{-10}}$ & 2.81 $\times$ ${10^{-11}}$ \\
      &o-\element[][]{H_2O} $3_{12}$ $\rightarrow$ $3_{03}$ & 7.84 $\times$ ${10^{-11}}$ & 3.20 $\times$ ${10^{-9}}$  & 1.68 $\times$ ${10^{-10}}$\\
\cline{2-5}
      &p-\element[][]{H_2O} $1_{11}$ $\rightarrow$ $0_{00}$ & 2.50 $\times$ ${10^{-7}}$  & 3.69 $\times$ ${10^{-6}}$  & 7.16 $\times$ ${10^{-7}}$ \\
      &p-\element[][]{H_2O} $2_{02}$ $\rightarrow$ $1_{11}$ & 2.70 $\times$ ${10^{-8}}$  & 8.08 $\times$ ${10^{-7}}$  & 5.83 $\times$ ${10^{-8}}$ \\
      &p-\element[][]{H_2O} $2_{20}$ $\rightarrow$ $1_{11}$ & 8.80 $\times$ ${10^{-10}}$ & 2.29 $\times$ ${10^{-8}}$  & 1.72 $\times$ ${10^{-9}}$\\
      &p-\element[][]{H_2O} $2_{11}$ $\rightarrow$ $2_{02}$ & 2.69 $\times$ ${10^{-10}}$ & 7.80 $\times$ ${10^{-9}}$  & 4.39 $\times$ ${10^{-10}}$ \\
\hline\hline
PDR II&o-\element[][]{H_2O} $1_{10}$ $\rightarrow$ $1_{01}$ & 7.25 $\times$ ${10^{-8}}$  & 6.72 $\times$ ${10^{-7}}$  & 1.55 $\times$ ${10^{-7}}$\\
      &o-\element[][]{H_2O} $2_{12}$ $\rightarrow$ $1_{01}$ & 7.30 $\times$ ${10^{-7}}$  & 5.71 $\times$ ${10^{-6}}$  & 1.81 $\times$ ${10^{-6}}$\\
      &o-\element[][]{H_2O} $2_{21}$ $\rightarrow$ $1_{10}$ & 1.79 $\times$ ${10^{-9}}$  & 2.59 $\times$ ${10^{-8}}$  & 3.09 $\times$ ${10^{-9}}$\\
      &o-\element[][]{H_2O} $2_{21}$ $\rightarrow$ $2_{12}$ & 1.27 $\times$ ${10^{-10}}$ & 1.85 $\times$ ${10^{-9}}$  & 2.18 $\times$ ${10^{-10}}$\\
      &o-\element[][]{H_2O} $3_{03}$ $\rightarrow$ $2_{12}$ & 1.48 $\times$ ${10^{-9}}$  & 2.24 $\times$ ${10^{-8}}$  & 2.61 $\times$ ${10^{-9}}$\\
      &o-\element[][]{H_2O} $3_{12}$ $\rightarrow$ $2_{21}$ & 4.28 $\times$ ${10^{-12}}$ & 7.84 $\times$ ${10^{-9}}$  & 1.06 $\times$ ${10^{-11}}$\\
      &o-\element[][]{H_2O} $3_{12}$ $\rightarrow$ $3_{03}$ & 2.54 $\times$ ${10^{-11}}$ & 4.65 $\times$ ${10^{-10}}$ & 6.34 $\times$ ${10^{-11}}$\\
\cline{2-5}
      &p-\element[][]{H_2O} $1_{11}$ $\rightarrow$ $0_{00}$ & 2.52 $\times$ ${10^{-7}}$  & 2.47 $\times$ ${10^{-6}}$  & 8.64 $\times$ ${10^{-7}}$ \\
      &p-\element[][]{H_2O} $2_{02}$ $\rightarrow$ $1_{11}$ & 7.72 $\times$ ${10^{-9}}$  & 1.50 $\times$ ${10^{-7}}$  & 1.72 $\times$ ${10^{-7}}$ \\
      &p-\element[][]{H_2O} $2_{20}$ $\rightarrow$ $1_{11}$ & 3.53 $\times$ ${10^{-10}}$ & 5.93 $\times$ ${10^{-9}}$  & 7.84 $\times$ ${10^{-10}}$ \\
      &p-\element[][]{H_2O} $2_{11}$ $\rightarrow$ $2_{02}$ & 8.35 $\times$ ${10^{-11}}$ & 1.51 $\times$ ${10^{-9}}$  & 1.81 $\times$ ${10^{-10}}$ \\
\hline\hline
PDR III&o-\element[][]{H_2O} $1_{10}$ $\rightarrow$ $1_{01}$ & 2.66 $\times$ ${10^{-8}}$  & 2.17 $\times$ ${10^{-7}}$  & 1.07 $\times$ ${10^{-7}}$\\
       &o-\element[][]{H_2O} $2_{12}$ $\rightarrow$ $1_{01}$ & 3.94 $\times$ ${10^{-7}}$  & 2.95 $\times$ ${10^{-6}}$  & 1.58 $\times$ ${10^{-6}}$\\ 
       &o-\element[][]{H_2O} $2_{21}$ $\rightarrow$ $1_{10}$ & 1.97 $\times$ ${10^{-10}}$ & 1.60 $\times$ ${10^{-9}}$  & 5.93 $\times$ ${10^{-10}}$\\
       &o-\element[][]{H_2O} $2_{21}$ $\rightarrow$ $2_{12}$ & 1.41 $\times$ ${10^{-11}}$ & 1.14 $\times$ ${10^{-10}}$ & 4.23 $\times$ ${10^{-11}}$ \\
       &o-\element[][]{H_2O} $3_{03}$ $\rightarrow$ $2_{12}$ & 8.81 $\times$ ${10^{-11}}$ & 3.82 $\times$ ${10^{-10}}$ & 1.04 $\times$ ${10^{-10}}$ \\
       &o-\element[][]{H_2O} $3_{12}$ $\rightarrow$ $2_{21}$ & 2.06 $\times$ ${10^{-13}}$ & 1.54 $\times$ ${10^{-12}}$ & 5.31 $\times$ ${10^{-13}}$ \\ 
       &o-\element[][]{H_2O} $3_{12}$ $\rightarrow$ $3_{03}$ & 1.23 $\times$ ${10^{-12}}$ & 9.12 $\times$ ${10^{-12}}$ & 2.68 $\times$ ${10^{-12}}$\\ 
\cline{2-5}
       &p-\element[][]{H_2O} $1_{11}$ $\rightarrow$ $0_{00}$ & 1.48 $\times$ ${10^{-7}}$  & 1.41 $\times$ ${10^{-6}}$   & 6.88 $\times$ ${10^{-7}}$ \\
       &p-\element[][]{H_2O} $2_{02}$ $\rightarrow$ $1_{11}$ & 4.55 $\times$ ${10^{-10}}$ & 3.95 $\times$ ${10^{-9}}$  & 1.08 $\times$ ${10^{-9}}$ \\
       &p-\element[][]{H_2O} $2_{20}$ $\rightarrow$ $1_{11}$ & 6.93 $\times$ ${10^{-11}}$ & 9.51 $\times$ ${10^{-10}}$ & 2.75 $\times$ ${10^{-10}}$ \\
       &p-\element[][]{H_2O} $2_{11}$ $\rightarrow$ $2_{02}$ & 4.81 $\times$ ${10^{-12}}$ & 2.72 $\times$ ${10^{-11}}$ & 7.39 $\times$ ${10^{-12}}$ \\
\hline
\end{tabular}
\end{minipage}
\end{table*}

\begin{table}
\caption{Inhomogeneous EMC model}
\begin{tabular}{lcc}
\hline  
Model & transition &average \\
 & &intensity \\
 & & \\
 & &$[$\mbox{$\mathrm{erg}\ \mathrm{cm}^{-2}\ \mathrm{s}^{-1}\ \mathrm{sr}^{-1}$}$]$ \\
\hline\hline
EMC core &o-\element[][]{H_2O} $1_{10}$ $\rightarrow$ $1_{01}$ & 1.37 $\times$ ${10^{-11}}$\\
               &o-\element[][]{H_2O} $2_{12}$ $\rightarrow$ $1_{01}$ & 5.45 $\times$ ${10^{-11}}$\\ 
               &o-\element[][]{H_2O} $2_{21}$ $\rightarrow$ $1_{10}$ & 2.78 $\times$ ${10^{-14}}$\\
               &o-\element[][]{H_2O} $2_{21}$ $\rightarrow$ $2_{12}$ & 1.98 $\times$ ${10^{-15}}$\\
               &o-\element[][]{H_2O} $3_{03}$ $\rightarrow$ $2_{12}$ & 2.90 $\times$ ${10^{-14}}$\\
               &o-\element[][]{H_2O} $3_{12}$ $\rightarrow$ $2_{21}$ & 1.53 $\times$ ${10^{-16}}$\\
               &o-\element[][]{H_2O} $3_{12}$ $\rightarrow$ $3_{03}$ & 9.17 $\times$ ${10^{-16}}$\\
\cline{2-3}
               &p-\element[][]{H_2O} $1_{11}$ $\rightarrow$ $0_{00}$ & 2.49 $\times$ ${10^{-11}}$\\
               &p-\element[][]{H_2O} $2_{02}$ $\rightarrow$ $1_{11}$ & 6.04 $\times$ ${10^{-14}}$\\
               &p-\element[][]{H_2O} $2_{20}$ $\rightarrow$ $1_{11}$ & 5.19 $\times$ ${10^{-15}}$\\
               &p-\element[][]{H_2O} $2_{11}$ $\rightarrow$ $2_{02}$ & 2.23 $\times$ ${10^{-15}}$\\
\hline
\end{tabular}
\label{tab:EMC}
\end{table}

\begin{table}
\caption{Antenna temperatures of the o-\element[][]{H_2O} $1_{10}$ $\rightarrow$ $1_{01}$ transition}
\label{tab:IAT}
\begin{tabular}{lc}
\hline  
Model &  antenna temperature  \\
 & $T_\mathrm{A}^*$\\
 & $[$ \mbox{K} $]$ \\
\hline\hline
Homogeen &  0.08\\ 
PDR I    &  0.25\\ 
PDR II   &  0.17\\  
PDR III  &  0.07\\  
EMC core &  3.6 $\times$ $10^{-4}$\\
EMC      &  3 $\times$ $10^{-4}$--3 $\times$ $10^{-3}$\\
\hline
\end{tabular}
\end{table}

The inhomogeneous nature of molecular clouds is well established from observations of extended $[$\ion{C}{i}$]$ and $[$\ion{C}{ii}$]$ (e.g., Keene et al. 1985; Plume et al. 1994) and has been modeled by various groups (e.g., Meixner $\&$ Tielens 1993; Spaans 1996; Stoerzer et al. 1996). It is therefore natural to explore the excitation of \element[][]{H_2O} in models incorporating high density clumps embedded in a low density interclump medium.\\ 
In the case of an inhomogeneous distribution, the radiative transfer is solved in three dimensions on a 81 $\times$ 81 $\times$ 81 grid.
In this section we not only cover the excitation of water in the PDR but also in the EMC, in order to fully interpret the results. We define the molecular cloud core as that part of the EMC that encloses the PDR and lies in one SWAS beam. What is left outside the molecular cloud core is denoted as the EMC.\\
The clumpy models have three free parameters: (i) the volume filling factor $F$, which gives the fraction of the total volume that is occupied by the clumps, (ii) a clump size $l_\mathrm{c}$, which fixes the extinction through an individual clump, and (iii) the clump-interclump density ratio $R_\mathrm{ic}$. For the PDR a fixed incident radiation field has been taken $I_\mathrm{UV}$ = 140 with respect to the Draine (1978) field, as well as a mean total hydrogen density $<$$n_\mathrm{H}$$>$ = 2 $\times$ $10^4$ \mbox{$\mathrm{cm}^{-3}$}, as in the case of the homogeneous PDR. The mean total hydrogen density $<$$n_\mathrm{H}$$>$ obeys $<$$n_\mathrm{H}$$>$ = $F$$n_\mathrm{c}$ + (1 - $F$)$n_\mathrm{ic}$. Here $n_\mathrm{c}$ and $n_\mathrm{ic}$ indicate the clump and interclump density, respectively. The enhancement factor for the EMC decreases to  $I_\mathrm{UV}$ $\approx$ 30--50 because of geometric dilution. The mean total hydrogen density $<$$n_\mathrm{H}$$>$ is taken to be $10^3$ \mbox{$\mathrm{cm}^{-3}$} over a region 10 \mbox{pc} in extent. The average hydrogen density of the EMC is not well constrained by observations, but the adopted value approximates those derived from the total \element[][13]{CO} column density divided by the size of the cloud (Blair et al. 1978; Plume et al. 1994).\\
Table \ref{tab:parameters} gives the parameters used for the different models. Three models have been run for the PDR and one for the EMC. The PDR clump densities range from 3$\times$${10^4}$ \mbox{$\mathrm{cm}^{-3}$} to 2$\times$${10^5}$ \mbox{$\mathrm{cm}^{-3}$}, whereas interclump densities vary between 3$\times$${10^3}$ \mbox{$\mathrm{cm}^{-3}$} and 5$\times$${10^3}$ \mbox{$\mathrm{cm}^{-3}$}. Different volume filling factors were incorporated, ranging from 2.5 $\%$ to 20 $\%$. In the PDR models, the clump size has been taken constant, $l_\mathrm{c}$ = 0.03 \mbox{pc}. The clump size in the EMC is taken to be 20 times larger than the clump size in the PDR, equal in proportion to the ratio in scale size. The velocity dispersion is taken to be 1.2 \mbox{$\mathrm{km}\ \mathrm{s}^{-1}$} in the PDR (Zhou et al. 1994), and 0.7 \mbox{$\mathrm{km}\ \mathrm{s}^{-1}$} in the EMC, assuming a weaker turbulent velocity component in comparison with the $\mathrm{v}_\mathrm{turb}$ of the PDR. The larger PDR velocity width simply reflects our assumption that the embedded sources inject some additional turbulent motions (IRS1--3, Blair et al. 1978; Beichman et al. 1979; Evans et al. 1989). This seems reasonable given that star-forming molecular cores tend to have larger (non-thermal) velocity widths than quiescent cores.\\
The clumps are distributed randomly in the interclump medium. The clump and interclump gas join smoothly with a density gradient. Due to the clumpiness the radiative transfer will differ from that of the homogeneous model. The UV radiation penetrates deeper into a clumpy region than into a cloud in which the gas is homogeneously distributed, leading to a lower water abundance because of enhanced photodissociation (Spaans $\&$ van Dishoeck 2001). The average water abundance in comparison with the homogeneous model drops by a factor $\sim$ 10. However, the resulting intensities depend sensitively on the optical depth, density and temperature structure along the line of sight. The inhomogeneity also affects the temperature structure, resulting in enhanced temperatures deeper into the interclump medium of the cloud, and lower clump temperatures. The result is an increase in the volume averaged gas temperature ($\sim$ 78 \mbox{K} for PDR I), compared with that of the homogeneous PDR model ($\sim$ 52 \mbox{K}). However, the mass-weighted average gas temperatures in the inhomogeneous models are 88 \mbox{K}, 70 \mbox{K} and 46 \mbox{K} for PDR I, PDR II and PDR III, respectively, and 52 \mbox{K} for the homogeneous model. This means that most of the mass of the cloud is located in cold clumps.\\
Intensities are computed in the same way as in the homogeneous model using Eq. (\ref{eq:Intensity}) and (\ref{eq:cooling}). We calculate the total intensity along each line of sight. Average and peak intensities for various transitions for the PDR are shown in Table \ref{tab:PDR}, and for the molecular cloud core, i.e., that part of the EMC lying at the southwest of the cloud which encloses the PDR, in Table \ref{tab:EMC}. Fig. \ref{fig:3dinhomorthodistr} and \ref{fig:3dinhomparadistr} show maps of the predicted distribution of the intensities for various transitions, and Fig. \ref{fig:WEcutinten} shows the change in intensity along a W-E cut for two ortho-\element[][]{H_2O} transitions. Note that the vast majority of the EMC is not contained in the SWAS beam. Snell et al. (2000) mapped a larger region of \object{S140} with SWAS, including several positions on the molecular cloud. They find no significant 557 \mbox{GHz} water emission residing in these areas (e.g., outside the molecular cloud core), in agreement with our results. The contribution of the EMC (core) to the total signal is negligible in comparison with the PDR. The minor contribution of the EMC (core) is due to the low gas density, which results in little excitation of the \element[][]{H_2O} molecule. Freeze-out of water is not required to explain the SWAS non-detections away from the PDR. However, the derived physical conditions indicate that freeze-out of water does occur.  From Bergin et al. (1995) we find a depletion factor of about 2--3 (see Section \ref{sec:sumanddisc}).\\
The intensity for the o-\element[][]{H_2O} ${1_{10}}$ $\rightarrow$ ${1_{01}}$ line in the homogeneous PDR model is a factor of a few lower than in PDR I and PDR II, and equals that of PDR III. The strongest transition we find is the o-\element[][]{H_2O} ${2_{12}}$ $\rightarrow$ ${1_{01}}$ in all models. This transition has an upper state energy of $\sim$ 80 \mbox{K} above ground level. The gas temperatures in the models are sufficient to excite this transition. All other transitions are a few orders of magnitude weaker than the two lowest transitions. To excite these transitions much higher temperatures are needed. For para-\element[][]{H_2O}, the lowest transition ${1_{11}}$ $\rightarrow$ ${0_{00}}$ is found to be the strongest in all the models.\\
From the average intensities, antenna temperatures are derived for the 557 GHz transition as seen in Table \ref{tab:IAT}, corrected for beam dilution in the case of the PDR with a beam dilution factor of 2. Note that the correct beam dilution factor in case of the PDR is uncertain. Since the antenna temperatures scale inversely in proportion to the beam dilution factor the results presented in Table \ref{tab:IAT} will differ when another beam dilution factor is incorporated.\\  
The total antenna temperature is the weighted sum of the PDR and EMC contributions. PDR I and PDR II in Table \ref{tab:PDR} yield, when combined with Table \ref{tab:EMC}, a value consistent with the observation of SWAS, i.e., $T_\mathrm{A}^*$ $\approx$ 0.25 \mbox{K}, (Ashby et al. 2000; Snell et al. 2000) for the lowest lying transition of ortho-\element[][]{H_2O}. At face value PDR I is in closest agreement with the observations.\\
Maps for the various transitions (line, not line + continuum) of the PDR I model are plotted in Figs. \ref{fig:3dinhomorthodistr} and \ref{fig:3dinhomparadistr}. It is seen that the o-\element[][]{H_2O} ${1_{10}}$ $\rightarrow$ ${1_{01}}$ ground state transition peaks near the eastern edge of the cloud. The bulk of the emission resides in the individual clumps. These clumps are low-temperature-water reservoirs. At the west side of the cloud the water abundance drops due to dissociation, still, for higher transitions, we see that the majority of the emission is shifted towards the western edge of the cloud. The \element[][]{H_2O} emission arises mainly from the warm, $\sim$ 60--110 \mbox{K}, clump edges, where the gas temperature is higher than in the interiors of the clumps. Conversely, the clump edge water abundance is a factor of $\sim$ 10 lower compared to the insides of the clumps, but still a factor of $\sim$ 10 higher than in the interclump medium. The 557 \mbox{GHz} line traces mostly the total column of water, while it is evident from Table \ref{tab:PDR} that the higher excitation lines differ from one model to the other more strongly.\\
Again, HIFI will be able to confirm the predicted shift towards the western edge of the higher water transitions because of its good angular resolution. Also, HIFI can constrain the inhomogeneity of the medium through water line observations.

\section{\label{sec:sumanddisc}Summary and Discussion}
In this paper we have examined the influence of inhomogeneity on the excitation of water in molecular clouds. Inhomogeneity causes photons to penetrate much deeper into the cloud leading to enhanced photodissociation of water. This gives rise to a lower water abundance. However, calculated intensities of PDR I and PDR II are a factor a few higher for the case of a homogeneous distribution. PDR III is in closest agreement with the intensities of the homogeneous model. This is not surprising since the parameters of PDR III resemble the homogeneous model most. This leads to the conclusion that the major part of the water line intensity emanates from the warm high density clump edges.\\
The total antenna temperature, $T_\mathrm{A}^*$, is reproduced for the o-\element[][]{H_2O} ${1_{10}}$ $\rightarrow$ ${1_{01}}$ line, observed by SWAS. However, the incorporated beam dilution factor is uncertain. The total signal is dominated by the signal of the PDR.\\
Water line emission is found also outside of the molecular cloud core, but is a few orders of magnitude weaker than in the PDR, in agreement with observations (Snell et al., 2000). In this paper, freeze-out of water onto dust grains has not been taken into account. Bergin et al. (1995) probed the evolution of molecular clouds, including depletion of atoms and molecules onto grain surfaces in the temperature and density regime we are interested in. They find that 50--60$\%$ of the water vapour can be removed from the gas phase, when gas temperatures are 20--25 \mbox{K}, dust temperatures 10 \mbox{K} and $<$$n_\mathrm{H}$$>$ = $10^3$ \mbox{$\mathrm{cm^{-3}}$}, as in the case of the EMC. Consequently, depletion can result in an even lower intensity. Depletion of water molecules in the PDR will occur only inside the shielded clumps (not the irradiated clump edges). We find for $n_\mathrm{H}$ $\sim$ $10^5$ \mbox{$\mathrm{cm^{-3}}$}, $T_\mathrm{g}$ = 25 K and  $T_\mathrm{d}$ = 15 K that $\sim$ 90$\%$ of the water vapour is frozen out (Bergin et al. 1995). \\ 
Predictions have been made for the intensities and integrated antenna temperatures of various transitions that HIFI will probe in the future. It is found for \object{S140} that the $2_{12}$ $\rightarrow$ $1_{01}$ and $1_{11}$ $\rightarrow$ $0_{00}$ transitions are the strongest for ortho- and para-\element[][]{H_2O}, respectively, in all the models.

\begin{acknowledgements}
We would like to thank the anonymous referee for constructive remarks on this work. We would like to thank Matt Ashby for helpful discussions.
\end{acknowledgements}

\bibliographystyle{bibtex/aa}
\bibliography{artikelIfinalversion}

\end{document}